\documentclass[a4paper,fleqn,usenatbib]{mnras}

\usepackage{mathptmx}
\usepackage{txfonts}
\usepackage{longtable}
\usepackage[T1]{fontenc}
\usepackage{ae,aecompl}

\usepackage{natbib} 
\usepackage{lscape}
\usepackage{graphicx}	% Including figure files
\usepackage{amssymb}	% Extra maths symbols
\title[The orbital periods of novae]{Life after eruption VIII: The orbital periods of novae}
\author[I. Fuentes-Morales et al.]{I. Fuentes-Morales$^{1}$\thanks{E-mail:
irma.fuentes@uv.cl (IF), claus.tappert@uv.cl (CT), monica.zorotovic@uv.cl (MZ)
},  C. Tappert$^{1}$\footnotemark[1], M. Zorotovic$^{1}$\footnotemark[1], N. Vogt$^{1}$, E. C. Puebla$^{1}$ \newauthor M. R. Schreiber$^{2, 3}$,  A. Ederoclite$^{4}$ and L. Schmidtobreick$^{5}$  \\
\\
$^{1}$ Instituto de F\'isica y Astronom\'ia, Universidad de Valpara\'iso, Avda. Gran Breta\~na 1111, Valpara\'iso, Chile\\
$^2$Departamento de F\'isica, Universidad T\'ecnica Federico Santa Mar\'ia, Av. Espa\~{n}a 1680, Valpara\'iso, Chile \\
$^3$Millennium Nucleus for Planet Formation, NPF, Valpara{\'i}so, Chile\\
$^{4}$ Instituto de Astronomia, Geof\'isica e Ci\^encias Atmosf\'ericas (IAG), Universidade de S\~ao Paulo (USP), Rua do Mat\~ao 1226,\\ C. Universit\'aria, 05508-090, S\~ao Paulo, Brazil\\
$^{5}$ European Southern Observatory, Casilla 19001, Santiago 19, Chile\\
}

\date{Accepted 2020 November 04. Received 2020 November 04; in original form 2020 June 17}

\pubyear{2020}

\newcommand{\Msun}{M_{\odot}}
\newcommand{\Rsun}{R_{\odot}}

\newcommand{\gppr}{\stackrel{>}{\scriptstyle \sim}}
\newcommand{\gappr}{\raisebox{-0.4ex}{$\gppr$}}

\begin{document}
\label{firstpage}
\pagerange{\pageref{firstpage}--\pageref{lastpage}}
\maketitle

\begin{abstract}
\noindent
The impact of nova eruptions on the long-term evolution of Cataclysmic Variables (CVs) is one of the least understood and intensively discussed topics in the field. A crucial ingredient to improve with this would be to establish a large sample of post-novae with known properties, starting with the most easily accessible one, the orbital period. Here we report new orbital periods for six faint novae: X Cir (3.71~h), IL Nor (1.62~h), DY Pup (3.35~h), V363 Sgr (3.03~h), V2572 Sgr (3.75~h) and CQ Vel (2.7~h). We furthermore revise the periods for the old novae OY Ara, RS Car, V365 Car, V849 Oph, V728 Sco, WY Sge, XX Tau and RW UMi. Using these new data and critically reviewing the trustworthiness of reported orbital periods of old novae in the literature, we establish an updated period distribution. We employ a binary-star evolution code to calculate a theoretical period distribution using both an empirical and the classical prescription for consequential angular momentum loss. In comparison with the observational data we find that both models especially fail to reproduce the peak in the  3 -- 4~h range, suggesting that the angular momentum loss for CVs above the period gap is not totally understood.

\end{abstract}


%%%%%%%%%%%%%%%%%%%%%%%%%%%%%%%%%%%%%%%%%%%%%%%%%%

%%%%%%%%%%%%%%%%% BODY OF PAPER %%%%%%%%%%%%%%%%%%

\section{Introduction}
A nova eruption occurs in Cataclysmic Variable stars (CVs), which are close interacting binary systems composed of a donor, usually similar to a late-type main-sequence star, that fills its Roche lobe, transferring material to the white dwarf (WD) primary component. If the accumulated hydrogen onto low-luminosity WD reaches a critical value, a thermonuclear runaway (TNR) is triggered on the surface of the primary that ejects material into the interstellar medium. This process is known as a nova eruption and CVs that experienced such an event are called classical novae or post-novae. The binary is not destroyed by the nova eruption, allowing for the accretion process to start anew, which possibly occurs as early as within one or two years after the eruption \citep{retter98}. The typical length of this recurrence cycle is currently estimated to $\ge$10$^4$ yr \citep{shara2012, linda2015}. This is thus not to be confused with the class of recurrent novae, which have much shorter recurrence cycles and stellar configurations that usually differ significantly from the main bulk of CVs.
%\textbf{On the other hand, in systems with accretion onto hot and massive WDs with no TNR are known as Super Soft X-ray Sources (SSS), and they have been proposed as supernova Ia progenitors \citep{starrfield04-sss,wheeler2012-SN}}. %The most mass lost in the CV is consequence of this event.

It is still not clear whether the behaviour of the CV between two subsequent nova eruptions is largely defined by the eruption, e.g.~with the latter causing the CV to switch between different states of mass-transfer rate ($\dot{M}$), or whether the CV is mainly unaffected by the eruption, e.g.~a low $\dot{M}$ pre-nova would emerge as a low $\dot{M}$ post-nova, and likewise for high $\dot{M}$ systems. In the latter case, it would be the intrinsic properties of the CV that determine the length of the nova cycle, without interaction with the nova eruption itself.
\\The first of the above possibilities has been investigated in greater detail by \citet{shara86}, leading to the postulation of the Hibernation model. There, the irradiation of the secondary star by the post-eruption heated WD causes the former to drive a very high $\dot{M}$ for a certain amount of time that gradually decreases as the WD cools down. Ultimately, this is supposed to lead to a detachment of the secondary star from its Roche lobe, thus stopping the transfer of material, and the system entering  ``hibernation''. As a consequence, all post-novae should appear as high $\dot{M}$ CVs (so-called nova-likes) during the decades or centuries following an eruption, then undergo a gradual transition into a low-$\dot{M}$ state and a dwarf-nova behaviour. As of yet, there is no clear evidence in favour or against this scenario. 
The discovery of former dwarf nova V1213 Cen appearing to transition to a brighter state with a stable, high-luminosity disc after the nova eruption is in good agreement with what is predicted by hibernation \citep{mroz2016}. However, it should be noted that the last observations of that study, about seven years after the eruption, still show the system in decline from the eruption with a significant slope $>0.1$ mag/yr, so it may well be that the object on a comparatively short time-scale returns to its dwarf-nova state. This would the be similar to the case of V446 Her, that was found to show dwarf-nova-like variability already about 30 years after the nova eruption \citep{v446her-dn-2011}. In a study of pre and post-nova brightness of 30 novae, \citep{collazzi2009} found that, while some objects present an increased luminosity after the eruption, most do not. Furthermore, \citet{weight94nohibernation} found that $\dot{M}$ did not decline for at least 140 yr after the eruption contrary to what the hibernation model predicts.
A recent study of the long-term behaviour of post-novae \citep{lifevii} also concluded that any decrease in $\dot{M}$ must be at much longer time scales than $\sim$200 yr.

An alternative explanation for the luminous accretion discs in post novae was given by \citet{schreiber2000}. There, the ionized state of the disc is caused by the WD irradiating the accretion disc, and not by an increased $\dot{M}$ from the secondary star. Depending on the size of the affected area in the disc, this would leave some outer parts in the disc in a non-ionized state, thus explaining the so-called stunted outbursts observed in some post-novae \citep[e.g.][]{honeycutt98}. \citet{tappertIII} indeed found evidence for the presence of an optically thick inner disc in one such object. In the same line, \citet{schreiber2001} concluded that the irradiated disc by the hot post-erupted WD plays a crucial role on the evolution of post-novae, with the decline in brightness being a direct consequence of the decrease of irradiation of the disc due to the cooling of the WD rather than an effect of a decrease in $\dot{M}$ as was interpreted by \citet{duerbeck92}.

% indicating that the irradiated inner disc by the hot WD post-eruption raises the brightness of the disc, imitating an increase in $\dot{M}$, and the gradual cooling of the WD could be interpreted as a decrease in $\dot{M}$.

%la hot wd after eruption ioniza el disco interno, lo que observacionalmente se interpretaría como un aumento en Mdot dps de la erupcion, imitando a un aumento de mdot, y que gradualmente a medida que disminuye la t de wd, el disco disminuye su brillo imitando low mdot 
% King et al -->  ciclo de mdot sin necesidad de tener nova eruption. 
%arguments that the different states in which mass transfer is seen in CVs would be a consequence of a cyclical process attributed to the irradiation by the WD, inducing changes in the radius of the secondary and consequently, to variations in the mass transfer \citep{king95}. 

One possibility to investigate the validity of above scenarios is to compare the physical parameters of the post-novae with those of the overall CV population. Of those, the orbital period ($P_\mathrm{{orb}}$) is the most accessible one and also represents already a rough indicator of the state within the secular evolution of CVs \citep[e.g.,][]{knigge2011-evolCVs}.
A number of theoretical orbital period distributions of novae have been published \citep{diaz97, nelson2004, townsleyBildsten2005}. However, for a proper comparison with the observed distribution, the latter needs to be made out of a sample of statistically significant size. 
The main problem related in general to the study of the post-nova population is that these are mostly very faint objects, requiring a significant amount of time on large telescopes to study them. 
%%%%%%%%%%%%%%%%%esto puede ir en el paper de los espectros de recover de nova %%%%%%%%%%%%%%%%%%%%%%%%%%%%
%because a significant amount of time on large telescopes is needed to recover them, due to their brightness having declined to a faint stable quiescence value after the eruption. Furthermore, most reported Galactic novae are located in the disc, which makes it more difficult to find them, and even more difficult to obtain time-resolved observations in order to determine its orbital periods. Despite of those obstacles, the number of orbital period known has been increasing with the time: 
\citet{diaz97} made the first observational period distribution of old novae from a sample of 28 novae with  $P_\mathrm{{orb}}$ < 10 h. Analysing the influence of certain observational selection effects, they found that those parameters have a little effect on the shape of the period distribution. They also suggested a correlation between the nova explosion amplitude and the orbital period. \citet{warner2002} analysed the period distribution using 50 orbital periods he qualified as reliable, indicating a concentration to 3.3 h, and noting a similarity to a pile-up of magnetic CVs near this value. 
 \citet{townsleyBildsten2005} used that period distribution to show that their simulations are consistent with the idea that CVs evolve across the period gap. \citet{tappertIII} compared the period distribution of all CVs (data from \citealt{RK2003}, version 7.20, 2013) with 78 orbital periods of post-novae. They confirmed the concentration of novae at 3-5 h, in striking difference to the distribution of all CVs. This particular range is dominated by high mass-transfer systems \citep{jil2007a}, in contrast to the general CV population, which is dominated by low-mass transfer and systems with orbital periods $<$ 2 h \citep{pala2020}. These differences were predicted by \citet{townsleyBildsten2005} and likely reflect the shorter nova eruption recurrence times for high-mass transfer systems.

However, the period distribution of novae is both still undersampled in large parts of the period range, such that an addition of a comparatively low number of new periods has the potential to significantly change the shape of the distribution. Therefore, any comparison with the predicted distribution will suffer from large uncertainties. This is the more important, because, since the brightness of the post-nova is mainly determined by the brightness of the accretion disc, the observed period distribution is potentially biased towards bright, long-period systems with high mass-transfer rates and low nova eruption amplitudes. Thus, short-period low mass-transfer novae could still amount to a significant number, but are hidden, because they are intrinsically faint. The work by \citet{gansicke2009} shows that observations of faint CVs are crucial for our understanding of CV evolution and the use of the period distribution as a diagnostic tool, and this likely is also the case for novae. 

In this work we derive the orbital period for a number of faint post-novae, and to improve the precision of already established periods for mostly eclipsing systems that were included in \citet{lifevii}. Furthermore, the theoretically predicted period distribution of novae based on a binary population model is calculated and compared to the observational data.

% The photometric and spectroscopic observations are detailed in section \ref{sec:s2}. Results are given in section \ref{sec:results}. In section \ref{sec:Porb-distribution} the theoretically predicted period distribution of novae based on a binary population model is calculated and compared to the observational data. Finally, section \ref{sec:summary} presents a summary and conclusions.

%%%%%%%%%%%%%%%%%%%%%%%%%%%%%%%%%%%%%%%%%%%%%%%%%%%%%%%%%%%%%%%%%%%%%%%%%%%%%%%%%%%%%%%%%%%%%%%%%%%%%%%%%%%
\section[]{ Observations and data reduction} \label{sec:s2}

\subsection{ Photometric data}
\label{subsec:photom}

We obtained time-series photometry in the {\it V}-band in 2013, 2014 and 2015 using direct CCD imaging with a field of view of 8.85 arcmin square,  0.259 arcsec pixel scale and a 2$\times$2 binning at the 2.5-m du Pont telescope at Las Campanas Observatory, Chile. 
Alignment of the individual images for each field was performed by the \textsc{astroimagej} software \citep{astroimagej}. All fields were reduced by bias and flat field correction and instrumental magnitudes were calculated with aperture photometry using the \textsc{daophot} package from \textsc{iraf}. The aperture radius in each frame was adopted as the average of the full width at half maximum (FWHM) of the stellar point spread function (PSF) in a given frame. Differential magnitudes were calculated using comparison stars in the vicinity of the post-nova, within a radius of 400 pixels. In order to calibrate the instrumental magnitude, stars with known  {\it  V} magnitude were chosen to be compared with their tabulated  {\it  V} magnitudes either in the The Naval Observatory Merged Astrometric Dataset (NOMAD, \citealt{nomadcatalogue}) or in the GSC (The HST Guide Star Catalogue, version 2.3.2) catalogue. The calculated  {\it  V} magnitudes are presented in the log of observations (Table \ref{tab:t1}).

Further  {\it  V}-band data were obtained between August 2013 and August 2015  with A novel Double-Imaging CAMera (ANDICAM) placed at the 1.3-m telescope operated by the Small and Moderate Aperture Research Telescope System (SMARTS) consortium, at the Cerro Tololo Inter-American Observatory (CTIO), located in La Serena, Chile. The field of view was six arcmin square and we used a 2$\times$2 binning. These observations yielded differential photometry over a time range of two years with a time resolution of the order of 3 -- 5 days. For more details concerning these data see \citet[][ paper VII]{lifevii}. Hereafter we refer to these observations as ``CTIO set''. 

\subsection{Spectroscopic data}

Time-series spectroscopic data were collected from the following observing runs: with the ESO Faint Object Spectrograph and Camera \citep[EFOSC2,][]{efosc2} at the ESO New Technology Telescope (NTT) in La Silla, Chile, we obtained data in June/July 2011, May 2012 and 2013. The grism used was $\#$20 covering a wavelength range of 6040 -- 7140 {\AA} with a 1 arcsec slit, yielding a resolution of 3.7 {\AA}. In December 2018 and January 2019, additional data on XX Tau were obtained at the Very Large Telescope (VLT) using the FOcal Reducer/low dispersion Spectrograph 2 \citep[FORS2,][]{fors2} with the 1200R grism and a 0.7-arcsec slit, covering a wavelength range of 5750 -- 7319 {\AA} with a resolution of 2.14 {\AA}. Acquisition frames were taken with the edge filter $GG435$, thus no broad-band photometric magnitudes are available for this run.  
%%% the spectral resolution given by the FWHM measured in an arc line is 1.5$\AA$ and the spectral resolution measured from the formula l/dl yields 2.13 
The nova RW UMi was observed in June 2015 with the Gran Telescopio Canarias (GTC), installed in the Spanish Observatorio del Roque de los Muchachos of the Instituto de Astrof\'isica de Canarias, in the island of La Palma, using the Optical System for Imaging and low-Intermediate-Resolution Integrated Spectroscopy \citep[OSIRIS,][]{osirisGTC98}. The R2500R volume-phased holographic grating was employed, covering a wavelength range of 5575 -- 7685 {\AA}. A 0.6 arcsec slit yielded a spectral resolution of 2.5 {\AA}, measured as the FWHM of the night-sky spectral lines. No acquisition frames were available, thus no estimates can be given for the photometric brightness of the object at 
the time of the observations.

The reduction and calibration of the data was conducted with \textsc{iraf}. Reduction of the spectra consisted in bias and overscan subtraction and division by a flat field that had been normalized by fitting a cubic spline of high order. The cosmic rays removal was performed with the \textsc{lacos$\_$spec} task for \textsc{iraf} \citep{lacos_cr}. One-dimensional spectra were extracted with the {\sc apall} routine within the {\sc onedspec} package. Wavelength calibration was determined with $He$, $Ar$ and $Ne$ lamp for datasets. The spectra were normalized with respect to the continuum and corrected to heliocentric velocity with the \textsc{iraf}'s \textsc{rvcorrect} task.

\subsection{Periodicity search}
\label{rv}

\noindent
While CVs are known for the presence of strong emission lines in their spectra, among whose H$\alpha$ is usually the most prominent one, most post novae actually show comparatively weak emission lines \citep[e.g.][and references therein]{tappertIV}, indicative of an optically thick accretion disc and a high mass-transfer rate. Additionally, most spectroscopic targets of the present study proved to be rather faint ($V > 18.0$), and thus the best signal-to-noise values did not exceed 5 and 10 for the EFOSC2 and the FORS2 data, respectively. This, together with most lines being broad, asymmetric, and of variable shape, rendered the usual methods of fitting the line profile to measure its Doppler shift unsuccessful. Thus, the technique used by \citet{tappertIII} to measure the H$\alpha$ displacement was employed: First, each normalized spectrum was smoothed down to the effective spectral resolution of the instrument. Second, to account for potential imperfections related to the wavelength calibration, individual wavelength corrections were applied with respect to the $\lambda$6300.304~{\AA} [O {\sc i}] sky emission line. Subsequently, the average spectrum for each target was cross-correlated by eye to each individual spectrum by applying a positional shift and an intensity scale factor. The resulting displacement was recorded as the radial velocity shift.

The periodicity analysis in both light curves and radial velocities was performed with \textsc{peranso} \citep{peranso}, which allows to choose among different methods based on discrete Fourier transform algorithm. The Lomb-Scargle routine was used and the error was estimated as the frequency resolution in each campaign. i.e. $1/\Delta t$.

Radial velocities are fitted with a sinusoidal function as: 
\begin{equation}
  \mathrm{{v_r(t)}}=  \gamma~ +~ K~ \sin[2 \pi~(t-T_{0})/ P_\mathrm{{orb}}]
 \label{eq:fit-rv}
\end{equation}
\noindent
Where: $v_r(t)$ is the measured radial velocity at time {\it t}, {\it K} corresponds to the semi amplitude, $\gamma$ is the systemic velocity, $T_{0}$ is the chosen zero point and $P_\mathrm{{orb}}$ is the orbital period of the system.

\begin{table}
%\centering

%\resizebox{0.5\textwidth}{!}{\begin{minipage}{\textwidth}
 \caption{Log of observations. Above: time-series photometry. Bottom: time-series spectroscopy. {\it N} refers to the number of observations, $t_{\mathrm{exp}}$ is the exposure time in seconds, $\Delta t$ is the time covered by observation in hours. Last column contains the magnitude value and the bandpass which it was measured. For the CTIO data, see \citet{lifevii}.}
% \scalebox{0.97} {
 
\begin{tabular}{@{}clllll}
\hline
 \multicolumn{1}{c}{Object} &
 \multicolumn{1}{c}{Date } &
   \multicolumn{1}{c}{{\it N}} &
  \multicolumn{1}{c}{$t_{\mathrm{exp}}$} &
   \multicolumn{1}{c}{$\Delta$t } &
  \multicolumn{1}{c}{magnitude } \\
 \multicolumn{1}{c}{} &
  \multicolumn{1}{c}{} &
   \multicolumn{1}{c}{} &
  \multicolumn{1}{c}{} &
   \multicolumn{1}{c}{} &
    \multicolumn{1}{c}{} \\
\hline
X Cir      & 2015-05-19  & 209 & 60  & 7.93  & 18.77(32)$V$ \\
           & 2015-05-20  & 80  & 60  & 2.89  & 18.76(38)$V$ \\
           & 2015-05-21  & 56  & 60  & 1.99  & 18.82(33)$V$ \\
           & 2015-05-23  & 37  & 60  & 1.48  & 18.81(37)$V$ \\
           & 2015-07-10  & 66  & 90  & 3.33  & 19.04(23)$V$ \\
IL Nor     & 2015-05-20  & 100 & 60  & 3.77  & 18.73(07)$V$ \\
           & 2015-05-21  & 53  & 60  & 1.89  & 18.52(07)$V$ \\
           & 2015-05-22  & 39  & 180 & 2.74  & 18.48(04)$V$ \\
           & 2015-05-23  & 42  & 60  & 1.66  & 18.24(07)$V$ \\ 
DY Pup     & 2013-12-31  & 139 & 90  & 5.11  & 19.16(07)$V$ \\
           & 2014-01-01  & 47  & 60  & 1.66  & 19.14(07)$V$ \\
V2572 Sgr  & 2012-05-16  & 142 & 40  & 3.15  & 17.92(08)$V$ \\ 
           & 2015-05-20  & 79  & 60  & 2.88  & 17.65(14)$V$ \\
           & 2015-05-21  & 121 & 60  & 4.36  & 17.77(10)$V$ \\
           & 2015-05-22  & 167 & 60  & 6.15  & 17.73(19)$V$ \\
           & 2015-05-24  & 90  & 60  & 3.34  & 17.68(08)$V$ \\
           & 2015-07-11  & 52  & 60  & 2.78  & 17.81(10)$V$ \\
XX Tau     & 2013-12-28  & 128 & 60  & 4.62  & 19.11(08)$V$ \\
           & 2013-12-30  & 45  & 60  & 1.59  & 19.15(09)$V$ \\
           & 2013-12-31  & 37  & 60  & 1.31  & 19.11(08)$V$ \\
           & 2014-01-01  & 18  & 120 & 0.92  & 19.05(07)$V$ \\
CQ Vel     & 2013-12-28  & 55  & 120 & 2.88  & 19.13(12)$V$ \\
           & 2013-12-29  & 79  & 120 & 5.37  & 19.05(11)$V$ \\
           & 2013-12-30  & 115 & 120 & 6.00  & 19.00(08)$V$ \\
           & 2013-12-31  & 22  & 120 & 1.12  & 18.99(08)$V$ \\
           & 2014-01-01  & 99  & 120 & 5.17  & 19.11(07)$V$ \\
\hline
V2572 Sgr  & 2011-06-29  & 1   & 900 & 0.25  & 18.30(59)$R$\\
           & 2011-06-30  & 3   & 900 & 3.54  & 18.52(15)$R$ \\
           & 2011-07-01  & 8   & 900 & 9.39  & 18.46(05)$R$ \\
           & 2011-07-03  & 1   & 900 & 0.25  & 18.64(10)$R$ \\
XX Tau     & 2018-12-30  & 8   & 600 & 0.90  & --\\
           & 2018-12-31  & 6   & 600 & 1.17  & --\\
           & 2019-01-01  & 20  & 600 & 3.72  & --\\
           & 2019-02-10  & 4   & 600 & 0.91  & -- \\
           & 2019-02-12  & 4   & 600 & 0.91  & -- \\
RW UMi     & 2015-06-19  & 10  & 600 & 1.56 & -- \\
           & 2015-06-21  & 16  & 600 & 2.60 & -- \\
           & 2015-06-22  &  5  & 600 & 0.69 & -- \\           
CQ Vel     & 2012-03-25  & 2   & 900 & 0.70  & 19.57(04)$R$\\
           & 2012-03-26  & 9   & 900 & 3.39  & 19.36(04)$R$\\   
           & 2012-03-27  & 7   & 900 & 2.69  & 19.35(28)$R$\\

\hline
                        
\end{tabular}
%}
\label{tab:t1}
\end{table}
%%%%%%%%%%%%%%%%%%%%%%%%%%%%%%%%%%%%%%%%%%%%%%%%%%
%%%%%%%%%%%%%%%%%%%%%%%%%%%%%%%%%%%%%%%%%%%%%%%%%
\section{results}
\label{sec:results}

\subsection{RS Car (1895)}

\begin{figure}
% \centering  % this centres figure in column
  \resizebox{0.5\textwidth}{!}{\includegraphics[trim={13mm 11mm 0mm 0mm}, clip]{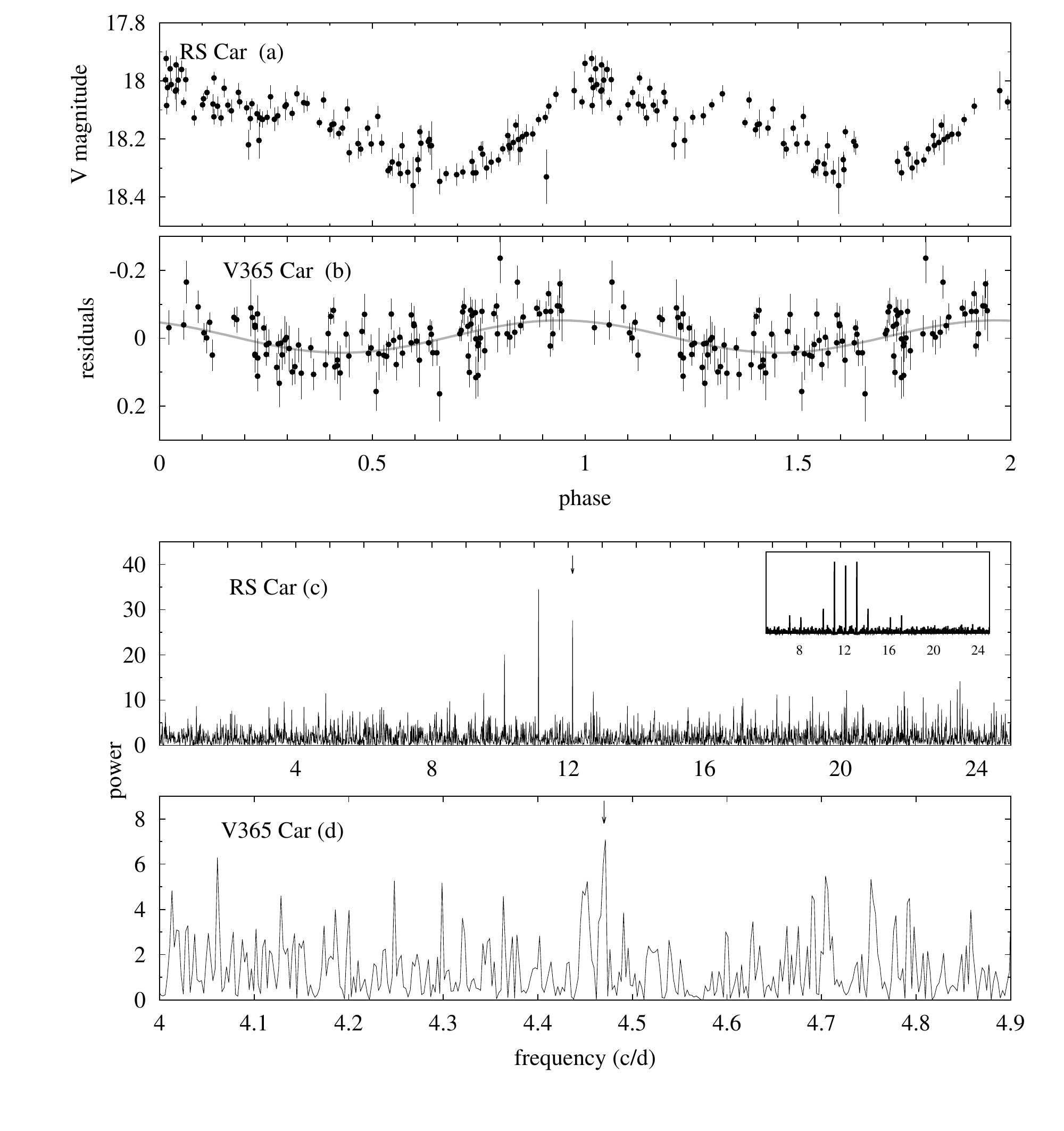}}
\caption{Phased light curves for (a) RS Car and (b) V365 Car according to ephemerides (\ref{eq:rscar-1}) and (\ref{eq:v365 car}) respectively. (c) Scargle periodogram for RS Car and the spectral window centered at the frequency marked with an arrow. (d) the same for V365 Car in the range of the frequency found by \citet{tappertIII}. The arrow marks the peak at {\it f} = 4.47 c/d. }
\label{fig:rscar-v365car} 
\end{figure}

This nova flared up in 1895 being discovered by Mrs. Fleming on photographic plates taken at the Arequipa Station of the Observatory \citep{rscar-discover1895}. The maximum light was reported at photographic magnitude $m_\mathrm{pg}$ = $\rmn{7.2^{m}}$. 
It was categorized as a slow nova and it was spectroscopically recovered by \citet{bianchini2001rscar} 7 arcsec away from the published position. The spectrum exhibited a blue continuum and a SED typically of an optically thick disc indicating that the system is still in a high-mass transfer state. \citet{woudt2002} presented high-speed photometry in white light of this nova, exhibiting a light curve with several features resembling strong flickering. While they do not present a plot of the Fourier spectrum, they describe it as consisting of a strong signal corresponding to $P = 1.977$ h,
i.e. 0.08238 d, and its harmonics. They ascribe this period to likely correspond to a superhump, based on RS Car showing the spectroscopic signatures of a high mass-transfer rate, which, at such short a period, is expected to produce an eccentric accretion disc, the latter being thought to be the physical reason behind the superhump signal \citep[e.g.,][]{wood2011-sh}. 
From our CTIO data, the periodogram presents two strong peaks at  ${\it f}_{1} = 11.13(01)$ and ${\it f}_2 = 12.13(01)$ c/d (Fig. \ref{fig:rscar-v365car}, c). The frequency resolution of the data set, 1/$\Delta t$, was used to estimate the associated uncertainty. We note that $f_2$ corresponds to a period that is very close to the signal detected by \citet{woudt2002}, implying that it is stable in time. We thus choose this as the main signal, in spite of it being the slightly lower of the two main peaks. Unfortunately, from the lack of corresponding information in \citet{warner2002}, we cannot examine the possible presence of $f_1$ in their data. 
Taking the maximum of the modulation according to ${\it f}_2$ as zero point, the ephemeris is
\begin{equation}
%RS Car    & h & 16676.7876(9) &  0.082429(3)    \\
\mathrm{HJD(max)} = 2\,456\,676.7876(09) + 0^{\mathrm d}.082429(25)~{\rmn{E}},
\label{eq:rscar-1}
\end{equation}
and the alternative ephemeris to ${\it f}_1$ is
\begin{equation}
\mathrm{HJD(max)} = 2\,456\,663.7723(16) + 0^{\mathrm d}.089842(81)~{\rmn{E}}.
\label{eq:rscar-2}
\end{equation}

\noindent
The phased light curves using the ephemeris (\ref{eq:rscar-1}) are shown in Fig. \ref{fig:rscar-v365car} (a). We note that the light curve shows similar characteristics as the one from \citet{woudt2002}, but the sequence of the humps has been inverted, with the large hump now following the minimum, and the small hump being the one preceding it. Other differences are that the minimum appears to be slightly broader (by about 0.1 phases) and that the total amplitude with $\sim$0.3$^\mathrm{m}$ is slightly larger.

\subsection{V365 Car (1948)}
\noindent
This nova with an eruption in 1948, discovered by  \citet{v365car-discover-henize67}, has been largely described by \citet{tappertIII}, who performed both radial velocities and {\it R}-band photometry. They found a periodicity to $P=$ 0.2247(40) d and their light curve present a sinusoid or hump shape with an amplitude of $\sim\rmn{0.2^{m}}$.   
The CTIO data on V365 Car showed a long-term decline in brightness \citep[see][for more details]{lifevii}. After subtracting this trend, we performed a period search on the residuals. While the resulting periodogram does not present any obvious dominant signal, a closer look at the frequency range near the previously reported value of {\it f} = 4.45 c/d by \citet{tappertIII} shows a narrow feature at {\it f} = 4.4704 c/d that rises slightly above the background noise (see Fig. \ref{fig:rscar-v365car}, d). Consequently, with the CTIO data was possible to refine the orbital period value.
The improved ephemeris of the maxima is
\begin{equation}
\mathrm{HJD (max)} = 2\,456\,628.845(78) + 0^{\mathrm d}.22369(12)~{\rmn{E}}.
\label{eq:v365 car}
\end{equation}
The folded light curve according to this period is shown in Fig.  \ref{fig:rscar-v365car} (b). Its shape as a sinusoid is similar to the one presented by \citet{tappertIII}, with an average amplitude of $\sim\rmn{0.1^{m}}$.   

%%%%%%%%%%%%%%%%%%%%%%%%%%%%%%%%%%%%%%%%%%%%%
\subsection{X Cir (1927)}
\label{subs:xcir}

\noindent
 X Cir underwent a nova eruption in 1927 \citep{xcirdiscover} and the position of the post-nova was recovered by \citet{tappertIV}. The spectrum indicated the presence of an accretion disc seen at high inclination, and the prominent Balmer emission lines along with a flat continuum point to low mass-transfer rate. 
 
\begin{figure}
\centerline{\includegraphics[width=.4\textwidth]{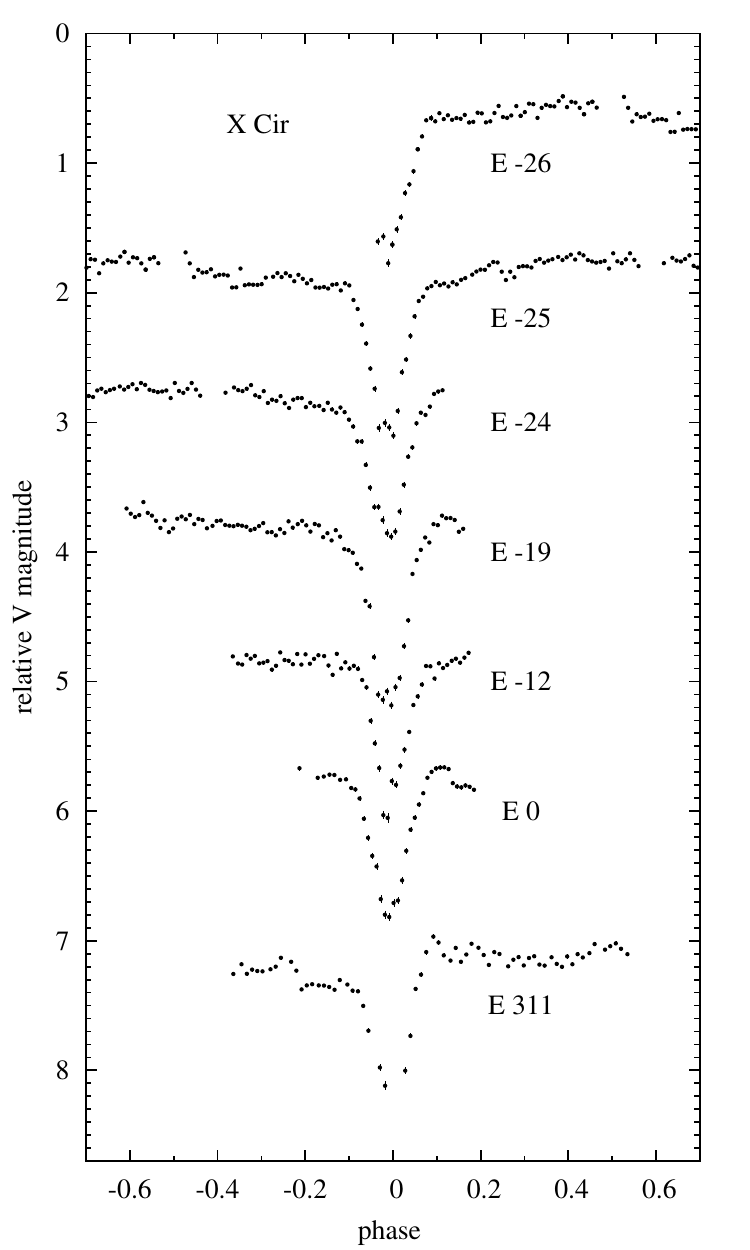}}
\caption{X Cir light curves from the du Pont telescope phased with the ephemeris (\ref{eq:xcir}). Each light curve was vertically shifted by 1.2 mag for the purpose of a clearer presentation.}
\label{fig:xCir-lc}
\end{figure}

 Special care was taken to perform the  {\it  V}-band photometry of this object, since a close visual companion is located at a distance of 0.8 arcsec southwestward. To assure a clean background subtraction and to account for the different seeing conditions, the aperture photometry was performed using a large annulus that covered both components of the visual binary. In good agreement with the conclusions drawn from the spectroscopic appearance, X Cir turned out to be an eclipsing CV with $P_\mathrm{{orb}}$ = 3.71\,h. The light curves are shown in Fig.~\ref{fig:xCir-lc}. A smooth variability is seen outside of the eclipse. The depth of the eclipse is slightly different in each cycle, varying from 1 mag to 1.5 over the seven observed cycles. At this stage it remains unclear whether these variations are intrinsic, or are caused by the presence of the companion in the aperture radius in combination with variable seeing.
 
X Cir was also part of the CTIO data set described in \citet{lifevii}, although it is not included in that paper, for reasons stated below. The data consist of 96 frames with typically two subsequent exposures per night with integration times of 170 and 340 s. The set spans a time range of 168.7 d, from HJD 2\,456\,690.7934 to 2\,456\,859.5209. Basic reduction was performed
as for the other objects of the CTIO data. However, because of the close companion, in combination with very variable seeing conditions, it was necessary to perform the aperture photometry of this object without applying a centering algorithm. For this purpose, one image frame with good seeing conditions was selected, and the positions of X Cir and the other component of the visual binary (hereafter M2) were measured with respect to a number of reference stars. In all other frames, the positions of those two components were calculated corresponding to the average of the shift of those reference stars with respect to the initially selected frame. Aperture photometry was performed at the such defined positions, and additionally of one comparison star whose constant brightness had been previously established. Finally, the differential magnitudes of the post-nova were computed as the difference between the brightness measured at its position and the average of the values of the comparison star and M2. 
The resulting data are shown in Fig.~\ref{fig:xcir-ctio} top. While it turned out that this light curve is still too strongly affected by the variable seeing to be used for a study of the intrinsic long-time behaviour of the post-nova, the fact that the data coverage includes a number of eclipses still made the set useful to refine above value of the orbital period obtained from the du Pont observations.

From the light curve, we identified 12 data points that could be unambiguously assigned to being part of an eclipse. Whenever there were two data points within the same night, we chose the fainter one as the time of eclipse, and in the cases where the two had identical brightness within the photometric uncertainty, we computed the average of those times. In order to calculate the correct cycles corresponding to each data point, we adjusted the orbital period iteratively. The value derived from the du Pont data was used to calculate the cycle corresponding to the second data point. A linear fit then yielded an improved period that was subsequently used to calculate the cycle corresponding to the third data point, and so forth. The fit to all six data points yielded $P_\mathrm{orb} = 0.154\,4504(38)$ d, which served to bridge the cycle count gap between the du Pont and the CTIO  data, and allowed for an unambiguous cycle count in the latter data set. The final fit to all eclipses gives the following ephemeris  

\begin{equation}
\mathrm{HJD(min)} = 2\,457\,166.5047(12) + 0^{\mathrm d}.154\,459\,53(63)~{\rmn{E}}~,
\label{eq:xcir}
\end{equation}
where we chose the cycle number of the best defined of the most recent eclipse measurements as zero point. The cycles, the measured eclipse times and the fit residuals are given in Table \ref{tab:t2}, and the CTIO phased light curve folded with this ephemeris is shown in the bottom plot of Fig. \ref{fig:xcir-ctio}. We ascribe the noisy eclipse shape and the light curve in general to the already mentioned different seeing conditions that caused a variable amount of the light of the close companion to be included in the aperture.

%%%%%%%%%%%%%%%%%%%%%%%%%%%%%%%%%%%%%%%%%%%%%%%%%%%%%%%%%%%%%%%%%%

\subsection{IL Nor (1893)}
\label{subs:ilnor}
\begin{figure}
 \resizebox{0.5\textwidth}{!}{\includegraphics[trim={10mm 25mm 0mm 0mm}, clip]{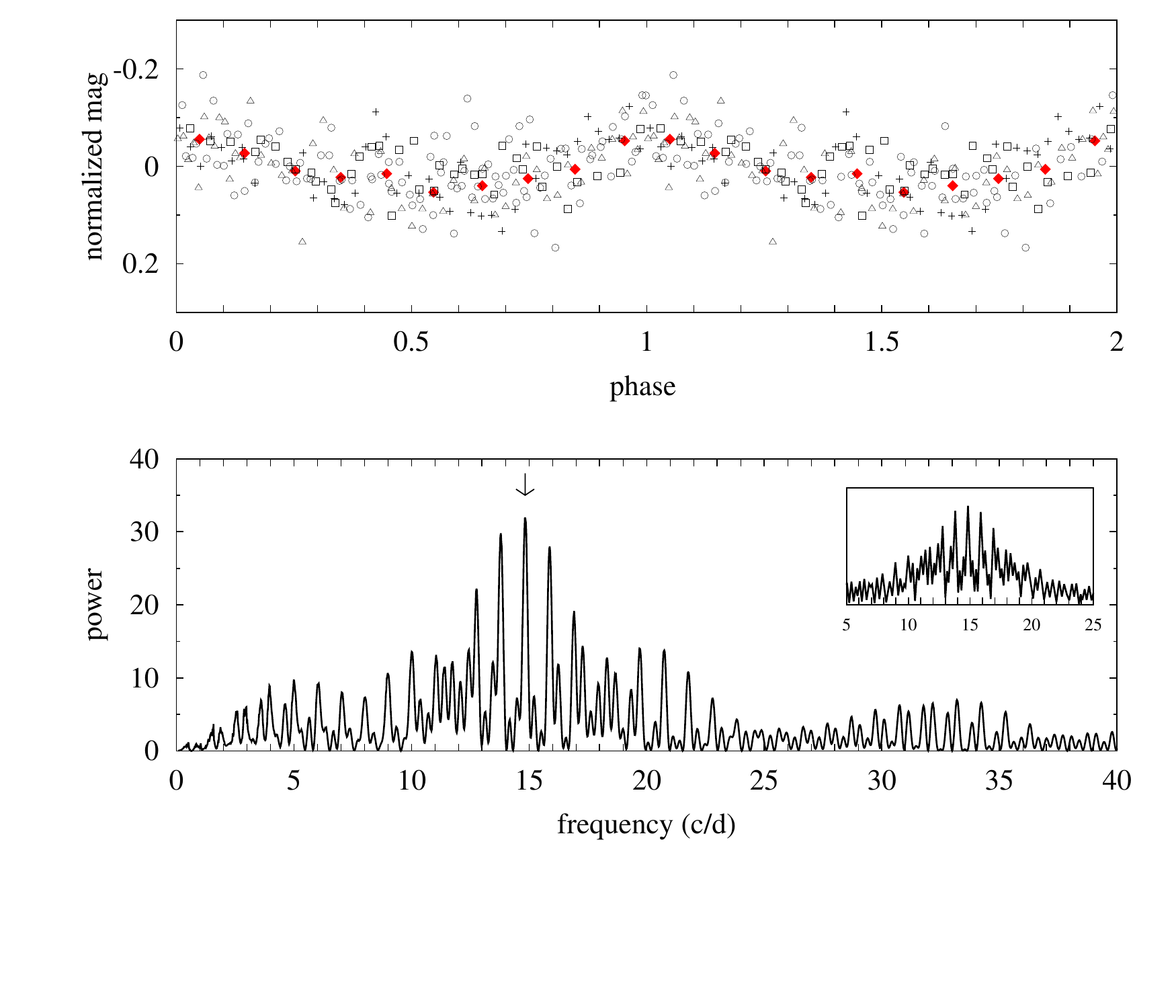}}
\caption{Top: Phased light curve for IL Nor according to ephemeris (\ref{eq:ilnor1}). The y axis corresponds to the normalized  {\it  V} magnitude and the x axis gives two orbital cycles in phase units. Different symbols indicate data from different nights. Red diamonds represent the average into 0.1 phase bins.  Bottom: Periodogram of the photometric data. The arrow marks the highest peak, corresponding to {\it f} = 14.83 c/d. The inset shows the spectral window.}
\label{fig:ilnor-phase}
\end{figure}

\noindent 
This is the oldest nova in the sample of new orbital periods, with an eruption reported in 1893 by Fleming and published by \citet{ilnordiscover}. It was identified by \citet{woudtwarner2010ilnor} based on photometric variability and spectroscopically confirmed by \citet{tappertI}. The spectrum is dominated by weak emission lines and a blue continuum, indicating equal to RS Car that, more than one hundred years after the eruption this object still is a high mass transfer rate system. 
Photometry made by us (see Fig.~\ref{fig:ilnor-lc}) at du Pont revealed strong short-term variability with an average {\it V} magnitude of $\rmn{18.5^{m}}$. In order to perform a period analysis, the {\it V} magnitude was normalized with respect to the mean of each night. The periodogram (Fig.~\ref{fig:ilnor-phase} bottom) shows a signal at $f_1$ = 14.83 c/d and strong aliases at  $f_2$ = 13.80 c/d and $f_3$ = 15.87 c/d which, if attributed to orbital modulation, correspond to $P_\mathrm{{orb1}}$ = 1.62(04) h, $P_\mathrm{{orb2}}$ = 1.74(04) h and $P_\mathrm{{orb3}}$ = 1.51(03) h respectively. Folding the data according to the alias frequencies does not present any significant differences with respect to the strongest peak. A comparison with the light curves of \citet{woudtwarner2010ilnor} does not resolve this ambiguity either. However, from the spectral window (Fig.~\ref{fig:ilnor-phase} bottom) it is evident that those peaks correspond to one cycle per day aliases. Considering the central peak in Fig. \ref{fig:ilnor-phase} at $f_1$, the ephemeris is:
 \begin{equation}
 {\rmn{HJD(max)}}=2\,457\,163.639(09) + 0^{\rmn{d}}.0674(15)~{\rmn{E}},
 \label{eq:ilnor1}
 \end{equation}
 
\noindent
and for $f_2$ and $f_3$ the ephemerides are
\begin{equation}
{\rmn{HJD(max)}}=2\,457\,163.643(07) + 0^{\rmn{d}}.0724(15)~{\rmn{E}}.
 \label{eq:ilnor2}   
\end{equation}
\begin{equation}
{\rmn{HJD(max)}}=2\,457\,163.642(05) + 0^{\rmn{d}}.0630(15)~{\rmn{E}},
 \label{eq:ilnor3}    
\end{equation}

\noindent
The phased light curve folded to ephemeris (\ref{eq:ilnor1}) together with its orbital phase averaged into 0.1 phase bins are shown in Fig. \ref{fig:ilnor-phase} (top) and could correspond to the orbital hump of IL Nor with a total amplitude $\sim \rmn{0.1^{m}}$. At least the two neighboring aliases mentioned above are also possible period solutions, requiring additional photometric and perhaps spectroscopic observations in order to decide which of the aliases is the valid one. In any case IL Nor turns out to be one of the very few classical novae below the period gap and is also the oldest confirmed nova among those short period systems.

%%%%%%%%%%%%%%%%%%%%%%%%%%%%%%%%%%%%%%%%%%%%%%%%%%%%%%%%%%%%%%%%%%

\subsection{DY Pup (1902)}
\label{subs:dypup}
\noindent
The nova eruption was discovered in November 19, 1902 on Harvard plates, being reported by \citet{shapley1921dypup}, who established the photographic magnitude at maximum  $m_\mathrm{pg}$ = $\rmn{7^{m}}$. He also found that the pre-nova had $m_\mathrm{pg}$ > $\rmn{10.3^{m}}$, and that must have been fainter than $\rmn{16^{m}}$ in 1901, because a photograph made in 1901 showing stars fainter than $\rmn{16^{m}}$ did not reveal any object at the nova position. DY Pup is catalogued as a slow nova considering the time it takes the brightness to decay by three magnitudes from maximum, i.e., $t_\mathrm{3}$ = 160 d  \citep{duerbeck87Catalogue}. %
The nova shell remnant is still visible and it was detected by \cite{Gill98Novashells} in 1995 as a ellipse-shaped remnant with a size of 7x5 arcsec. Despite its detection, the distance could not be estimated due to the lack of information on the expansion velocities.
Comparison of the finding chart in \citet{downes2005} and the images of the Panoramic Survey Telescope and Rapid Response System \citep[Pan-STARRS,][]{panstarrs2016,panstarrs1} unambiguously identifies DY Pup with a source in the Gaia Data Release 2 catalogue \citep{gaiaDR2}. However, the measured parallax is 0.26 $\pm$ 0.31 mas, and thus presents an uncertainty that is too large for a meaningful distance determination \citep{dist-parallax2015,gaiadistancesShaefer18, tappert20arxiv}.

\begin{figure}
% \centering  % this centres figure in column
  \resizebox{0.4\textwidth}{!}{\includegraphics[trim={0mm 0mm 0mm 0mm}, clip]{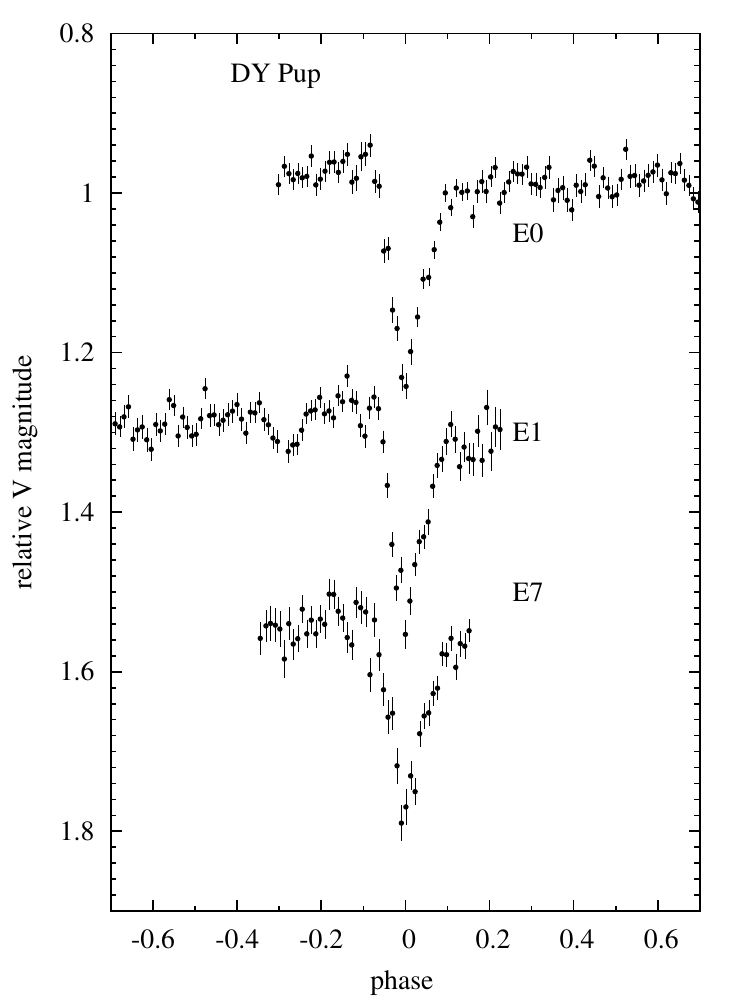}}
%\centerline{\includegraphics[scale=.9]{fase_DYPup_lc.pdf}}
\caption{ V magnitude vs phase using the ephemeris (\ref{eq:dypup}) for DY Pup.}
\label{fig:dypup}
\end{figure}

Only two spectral observations have been reported \citep{zwitter94,tomov2015dypup}. Both spectra are dominated by a blue continuum and weak H$\alpha$ emission line. In a poster presentation, and in a later proceeding, Van Zyl reported that DY Pup is an eclipsing system with $P_\mathrm{orb} = 3.35$ h \citep{downes2001, warner2003dypup}, but the corresponding light curves were not published. Our  {\it  V} photometric observations confirm this information, detecting three eclipses during our two nights of observations (Fig. \ref{fig:dypup}, Table \ref{tab:t2}). The corresponding ephemeris for the mid-eclipse timing results to
\begin{equation}
{\rmn{HJD(min)}}=2\,456\,658.64779(74) + 0^{{\rmn{d}}}.13952(25)~{\rmn{E}}.
\label{eq:dypup}
\end{equation}

\noindent
The eclipse is comparatively shallow, with a depth of $\sim\rmn{0.3^{m}}$. The very small amount of flickering in the light curve and the diminished pre-eclipse hump indicates that DY Pup is a high mass-transfer CV.

%%%%%%%%%%%%%%%%%%%%%%%%%%%%%%%%%%%%%%%%%%%%%%%%%%%%%%%%%%%%%%%%%%
\subsection{V363 Sgr (1927)}
\label{subs:v363sgr}

\begin{figure}
 \resizebox{0.5\textwidth}{!}{\includegraphics[trim={10mm 0mm 0mm 0mm}, clip]{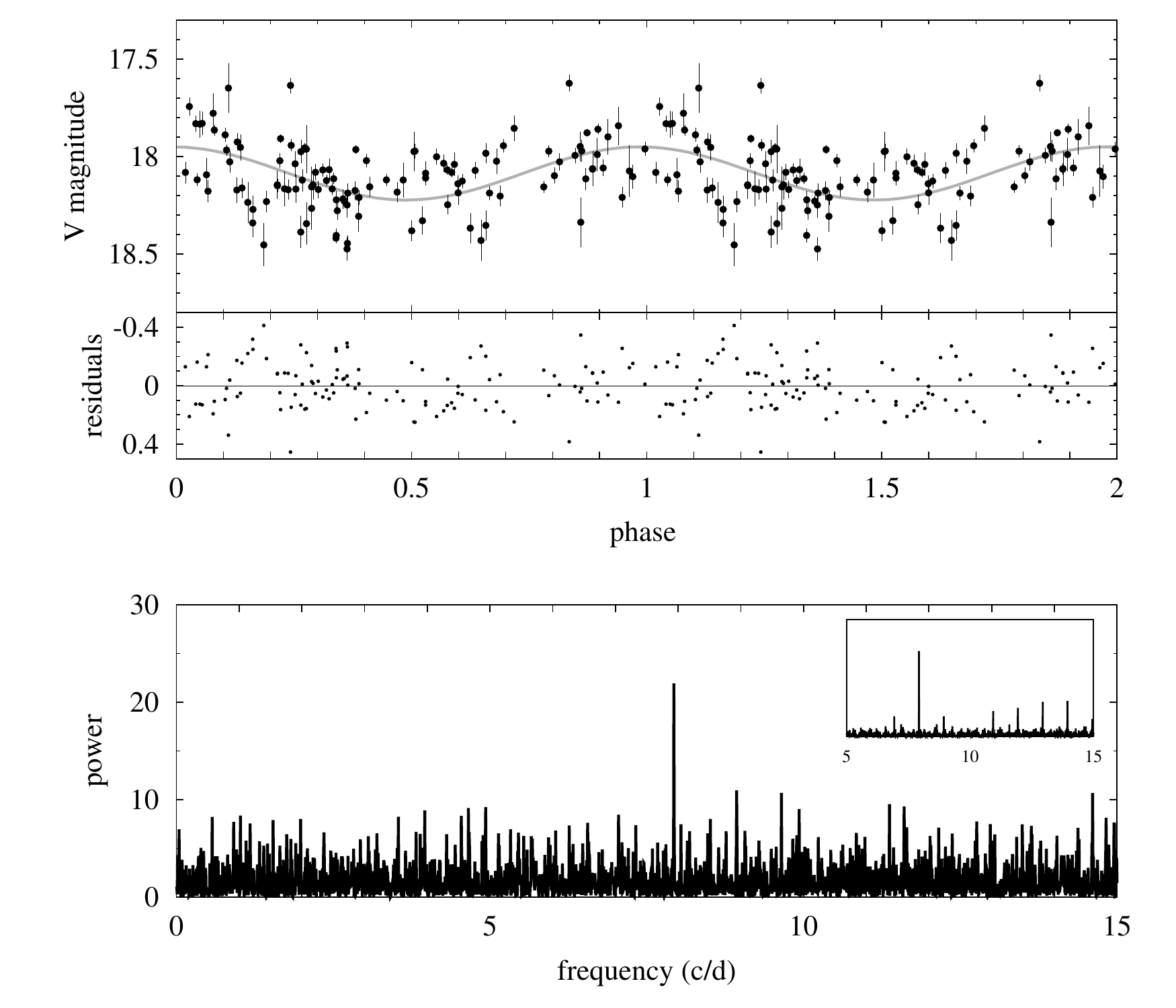}}
\caption{Top: Phased light curve for V363 Sgr according to ephemeris (\ref{eq:v363sgr}) together with a sine fit (grey line). Middle: Residuals of the fit. Bottom: Periodogram showing the highest peak corresponding to {\it f} = 7.93 c/d. As inset plot the spectral window centered at this frequency is shown.}
\label{fig:v363sgr}
\end{figure}

\noindent
For a long time, the identification of this post-nova was ambiguous. \cite{tappertIV} found, $\sim$40 arcsec from the published position, a star with a blue continuum and narrow and weak emission lines. They suggest a low orbital inclination, but a rather high accretion rate. No orbital period of this star has been published. This nova was part of our CTIO data set, consisting of typically two subsequent data points per night every three nights over a range of 356 d. A period analysis of that data revealed an unambiguous signal at {\it f} = 7.93 c/d that corresponds to a periodic hump or sinusoidal variation with $P = 3.03$ h which we interpret as the orbital period (Fig. \ref{fig:v363sgr}). The corresponding ephemeris for the maximum is 
 \begin{equation}
{\rmn{HJD(max)}}=2\,456\,583.579(45) + 0^{{\rmn{d}}}.126066(95)~{\rmn{E}}.
\label{eq:v363sgr}
\end{equation}

\noindent
This value places V363 Sgr inside the period gap of CVs as defined by \citet{knigge2006periodgap}. The existence of the photometric modulation indicates a medium-high inclination, somewhat contradicting the conclusion by \citet{tappertIV} based on the narrow emission lines. However, V363 Sgr could also be a permanent superhumper which allows for lower inclinations \citep{smak2010sh}. In this case the orbital period could be a few per cent different from the above value.

%%%%%%%%%%%%%%%%%%%%%%%%%%%%%%%%%%%%%%%%%%%%%%%%%%%%%%%%%%%%%%%%%%%%%%%%%

\subsection{V2572 Sgr (1969)}
\label{subs:v2572sgr}

\begin{figure}
 \resizebox{0.5\textwidth}{!}{\includegraphics[trim={10mm 1mm 0mm 0mm}, clip]{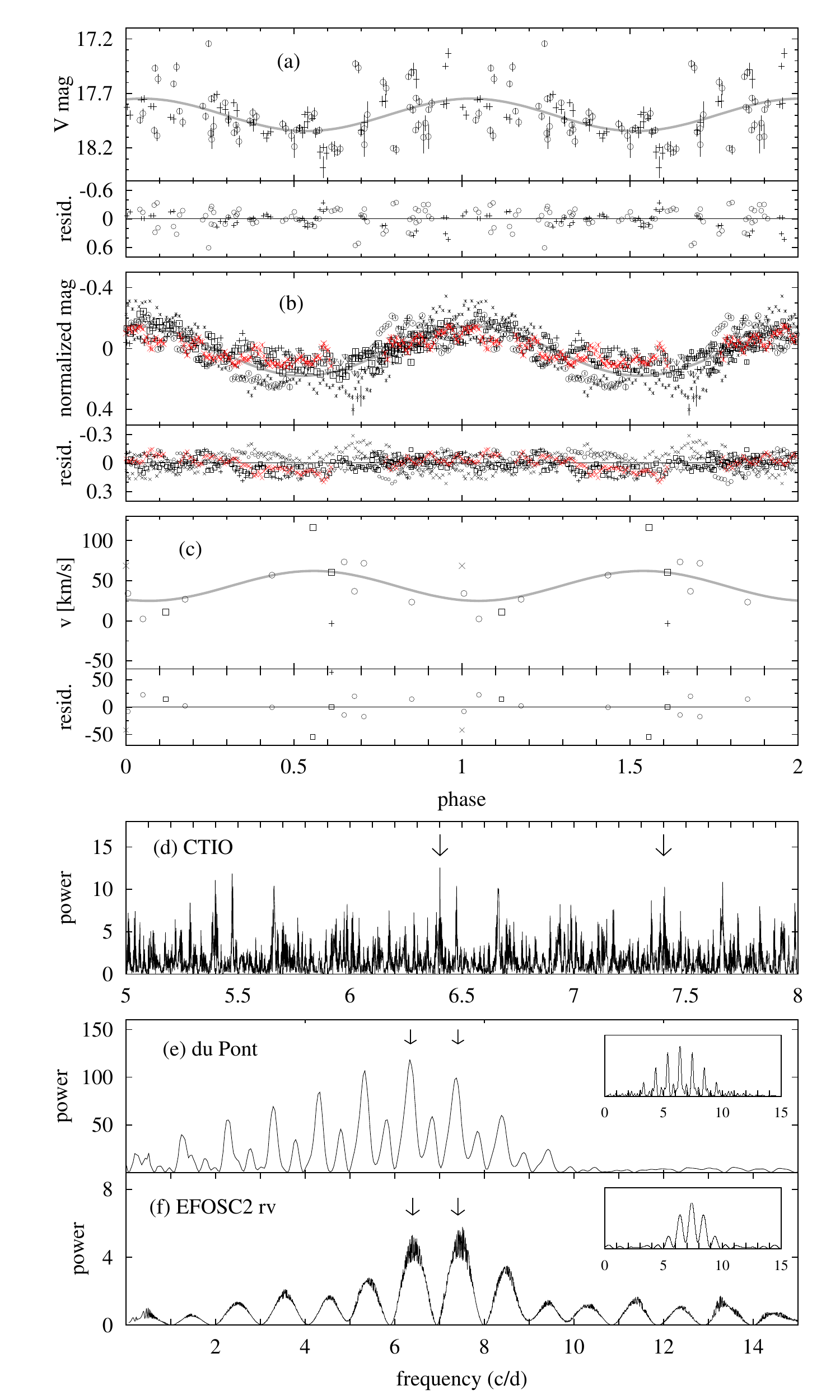}}
\caption{(a): Phase light curves of V2572 Sgr folded with the ephemeris (\ref{eq:v2572sgr-1}) for CTIO data. set 1 and 2 are shown with crosses and circles respectively. (b) The same for EFOSC2/NTT's light curve (red crosses) and du Pont observations  (different black symbols represent different nights). (c) Radial velocity fitted with the ephemeris (\ref{eq:v2572sgr-1}). A sine curve fitted to the data as a grey line and residuals are also shown. Scargle Periodogram for (d) CTIO data in the range of {\it f} = 5 -- 8 c/d. (e) du Pont photometric data (f) radial velocity data. As inset plot is shown the spectral window centered at the dominant frequency. The arrows point to {\it f} = 6.40 c/d  and {\it f} = 7.41 c/d (see text for details). }
\label{fig:v2572Sgr-phase}
\end{figure}

\noindent
 \cite{tappertI} give a brief description of the eruption light curve of this object and present a spectrum with comparatively weak Balmer emission lines, the He {\sc i} series and Bowen/He {\sc ii}. They concluded that V2572 Sgr could be a high mass transfer system. In our attempts to determine its period, the periodogram of our radial velocities measured with EFOSC2 in 2011 showed a strong and broad peak at {\it f} = 7.49(07) c/d  corresponding to $P_\mathrm{orb}$ = 3.20 h and a broad and predominant alias at {\it f} = 6.45(02) c/d (Fig. \ref{fig:v2572Sgr-phase}, f). One {\it V}-band light curve with 3.15 h time span, obtained with the same instrument in 2012 exhibited a hump structure with strong flickering (Fig. \ref{fig:v2572Sgr-lc}). If an orbital signature is present, the period should be larger than 3.15 h, because these data clearly do not cover a full orbit, thus frequencies $> 7.6$ c/d can be discarded. The {\it V}-band light curves taken at du Pont reveal a periodic hump with a variable amplitude, up to $\sim\rmn{0.3^{m}}$. The periodogram of this campaign (Fig. \ref{fig:v2572Sgr-phase}, e) shows a central peak at frequency 6.38(04) c/d and two aliases at 5.35(08) and 7.41(08) c/d of similar height as the central one, being these values comparable to those found in the radial velocities periodogram. Folding the radial velocities and the du Pont photometry according to these frequencies yielded reasonable light and radial velocity curves for the frequencies 6.38 and 7.41, but systematically offsets for individual data sets from the general behaviour for $f=5.38$ c/d, so that it was discarded.

V2572 Sgr was also included in the CTIO data set, in two seasons, implying a total coverage of nearly one year (see Fig. \ref{fig:v2572sgr-ctio}). The search for periodicities was performed independently in each of the two data sets as well as combining all the data to a single set. Its periodograms present large noise level due to the high cadence (only two points per night), however a narrow and outstanding frequency at $f_1$ = 6.40(01) c/d in each of the single sets and in the combined one is present, together with several frequencies of similar height (Fig. \ref{fig:v2572Sgr-phase}, d). The frequencies at 6.47(01) and 6.66(01) c/d were discarded because the light curves from du Pont are not fitted properly with these periods. Comparing with the results from the du Pont data, this leaves only two viable frequencies $f_1 = 6.40(01)$ and $f_2 = 7.41(01)$ c/d.  Accordingly, we assumed that the orbital frequency could be equivalent to $P_1 $ = 0.156211(29) d or $P_2 $ =  0.135125(22). For both, the rather accurate period allowed to the bridge the CTIO data to those of du Pont enabling to derive a unique cycle number difference between their epochs and, consequently, also to that of the early EFOSC2 run.  

For the final ephemerides we used not only the hump maxima, but also the minima which happen to appear always very near to phase 0.5 in all time-resolved data. Those epochs correspond to the extremes of a polynomial function of degree two fitted for both EFOSC2 and du Pont HJD data. For CTIO data those epochs were derived from the phase data plot. We identified the HJD of the points located close to zero phase and those close to phase 0.5. Making sure that a slight variation of the orbital period did not have a markedly effect on their position in phase space they were then counted as maxima and minima, i.e. the respective HJDs were assigned to full and half cycles, respectively. It should be mentioned that a high uncertainty is associated to this calculation, since those two points per night can correspond to any part of the wide hump. Table \ref{tab:humps} gives the resulting cycle numbers ${\rmn{E}}$ and ${\rmn{HJD}}$ epochs considering both periods; integer numbers ${\rmn{E}}$ refer to observed maxima, the remaining ones to minima. A least square fit through the data ${\rmn{E}_1}$ in Table \ref{tab:humps} yields the ephemeris for the hump maximum 
  \begin{equation}
 {\rmn{HJD(max)}}=2\,456\,507.6959(66) + 0^{{\rmn{d}}}.1562146(19)~{\rmn{E}},
 \label{eq:v2572sgr-1}
 \end{equation} 
with a standard deviation of $\sigma$ = 0.018 d and for $f_2$
  \begin{equation}
 {\rmn{HJD(max)}}=2\,456\,507.6563(66) + 0^{{\rmn{d}}}.1351221(16)~{\rmn{E}},
 \label{eq:v2572sgr-2}
 \end{equation} 
with a standard deviation of $\sigma$ = 0.018 d. Phased light curves considering the ephemeris (\ref{eq:v2572sgr-1}), for all photometric data sets are shown in the upper part of Fig. \ref{fig:v2572Sgr-phase}. The sinusoidal parameters for the radial velocities listed in Table \ref{tab:rv} according to $P_1$ are $\gamma$ = 44(7) km/s and $K$ = 19(8) km/s and for $P_2$ are $\gamma$ = 47(5) km/s and $K$ = 26(8) km/s.

 %%%%%%%%%%%%%radial velocities parameters\[Gamma]=43(7) K=19(8) 

%%%%%%%%%%%%%%%%%%%%%%%%%%%%%%%%%%%%%%%%%%%%%%%%%%%%%%%%
%%%%%%%%%%%%%%%%%%%%%%%%%%%%%%%%%%%%%%%%%%%%%%%%%%%%%%%%

\begin{table}
\centering

%\resizebox{0.5\textwidth}{!}{\begin{minipage}{\textwidth}
%\begin{minipage}{0.5\columnwidth}
 \caption{Epochs of humps observed in the time-resolved data for V2572 Sgr. The $\rmn{E_1}$, $\rmn{(O-C)_1}$  and $\mathrm{E_2}$, $\mathrm{(O-C)_2}$ values refer to Eq. (\ref{eq:v2572sgr-1}) and to Eq. (\ref{eq:v2572sgr-2}) respectively.}
\scalebox{0.9} {
\begin{tabular}{@{}lrrrr}
\hline

 \multicolumn{1}{c}{{$\rmn{HJD}$}} &
   \multicolumn{1}{c}{$\mathrm{E_1}$} &
  \multicolumn{1}{c}{$\mathrm{E_2}$} &
    \multicolumn{1}{c}{$\mathrm{(O-C)_1}$ }  &
 \multicolumn{1}{c}{$\mathrm{(O-C)_2}$ }   \\
  \multicolumn{1}{c}{ $-2\,450\,000$ d} &
 \multicolumn{1}{c}{} &
  \multicolumn{1}{c}{  } &
  \multicolumn{2}{c}{(d)}  \\

\hline
\multicolumn{5}{c|}{EFOSC2}\\
6063.759(06) & -2842  & -3285    & 0.023   & -0.021  \\        
6063.829(70) & 2841.5 & -3284.5  & 0.015   & -0.019   \\     
\hline                                                         \multicolumn{5}{c}{CTIO}\\              
6507.684(25) &    0   & 0        & -0.014  &  0.028  \\
6510.732(36) & 19.5   & 22.5     & -0.013  &  0.035  \\
6792.907(50) & 1826   & 2111     & -0.039  &  0.008   \\
6838.770(30) & 2119.5 & 2450.5   & -0.025  & -0.003  \\
\hline    
\multicolumn{5}{c}{du Pont}\\
7162.853(08) & 4194   & 4849     & -0.009  & -0.010  \\
7163.810(02) & 4200   & 4856     &  0.010  &  0.001  \\
7163.890(08) & 4200.5 & 4856.5   &  0.012  &  0.013  \\
7164.748(01) & 4206   & 4863     &  0.011  & -0.007  \\
7164.822(01) & 4206.5 & 4863.5   &  0.007  & -0.001  \\
7164.903(07) & 4207   & 4864     &  0.010  &  0.013  \\
7166.765(03) & 4219   & 4878     & -0.003  & -0.017  \\
7214.662(08) & 4525.5 & 5232.5   &  0.015  & -0.021  \\
\hline

\end{tabular}
}

%\end{minipage}
\label{tab:humps}
\end{table}

%%%%%%%%%%%%%%%%%%%%%%%%%%%%%%%%%%%%%%%%%%%%%%%%%%%%%%%%%%%%%
\subsection{XX Tau (1927)}
The history of this nova that erupted in 1927 has been extensively described by \citet{linda2005xxtau} who also present an optical spectrum dominated by strong Balmer and He{\sc i} emission lines resembling more a  dwarf nova than an old nova. However, the CTIO data did not present any clear evidence for outburst behaviour in roughly 1.5 yr spanning monitoring.  

\citet{rodriguez-gil2005} found a number of periodicities in time-series photometric data at periods of 23.69(03) min, 3.26(05) h and 5 d. While the shortest value was considered as very uncertain, the middle one was attributed to an orbital or superhump modulation, and the longest period was interpreted as evidence of an eccentric/tilted accretion disc. 

\begin{figure}
% \centering  % this centres figure in column
  \resizebox{0.5\textwidth}{!}{\includegraphics[trim={12mm 10mm 0mm 0mm}, clip]{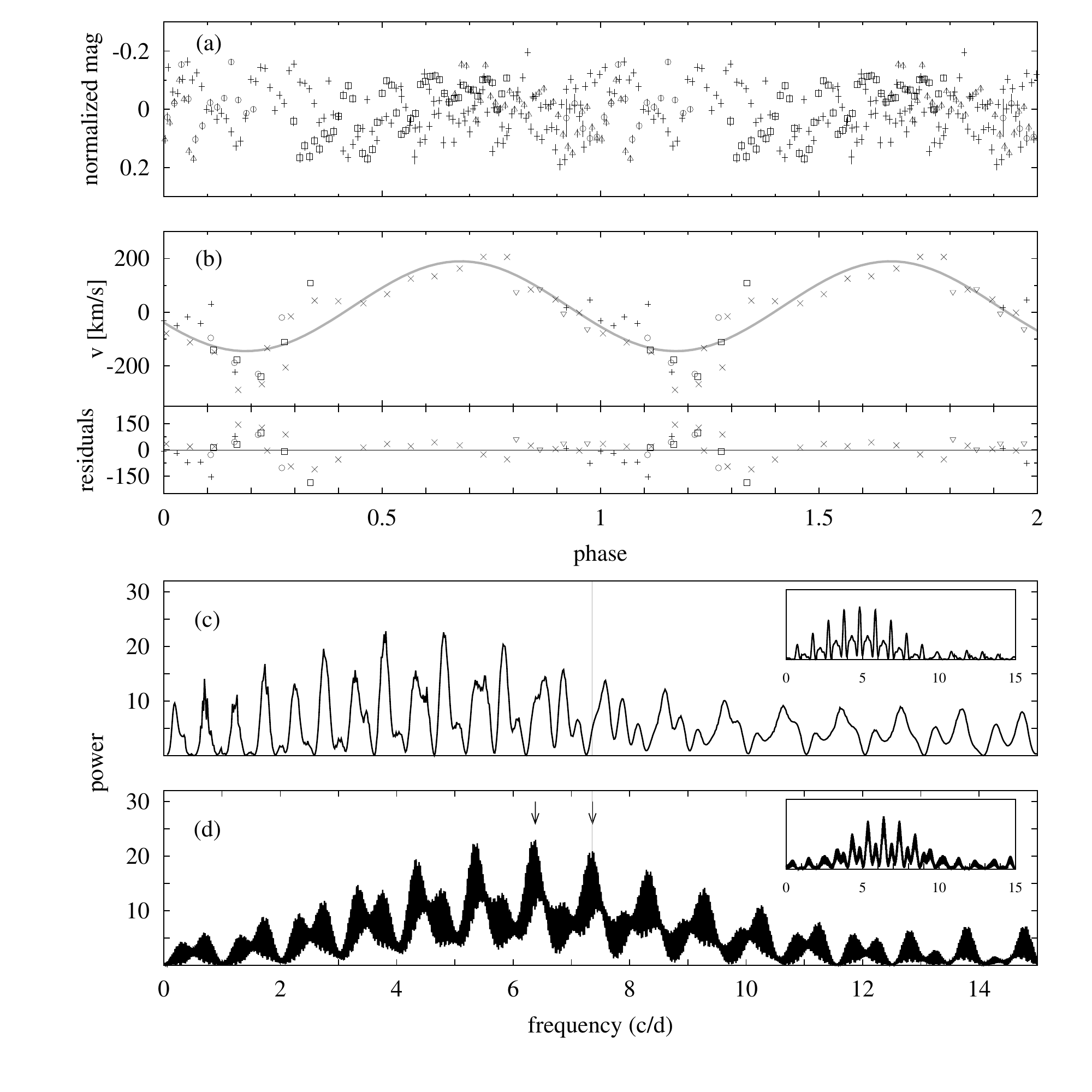}}
\caption{(a) Photometric phased light curve for XX Tau according to ephemeris (\ref{eq:xxtau2}). (b) Radial velocity, the sine fit and the residuals. Different nights are shown as different symbols. Phases very likely do no coincide because two data sets have different $T_0$. (c) Periodogram of the photometric data together with the spectral window centered at the main frequency. Vertical grey line indicates the position of the detection made by \citet{rodriguez-gil2005}. (d) Periodogram of the radial velocities. The arrows point to the frequencies $f_1$ and $f_3$ (see text for details).} 
\label{fig:xxtau}
\end{figure}

Our light curves taken in a five nights spanning observing run at the 2.5-m du Pont are dominated by strong irregular flickering. The periodogram does not show any sign of the suspected orbital or superhump modulation (Fig. \ref{fig:xxtau}, c). Instead, its highest peak is at $f$ = 4.82(06) c/d, equivalent to $P$ = 4.98 h. However, this signal is clearly not stable (Fig. \ref{fig:xxtau-lc}), and thus probably is simply caused by flickering mimicking a periodicity within our comparatively short time-series.

Radial velocities measured from time-series spectroscopic data taken at FORS2/VLT in five nights spanning two weeks in total show a periodogram with a central peak and a number of significant broader one-day aliases, each one
composed of a number of narrow peaks as can be seen in bottom panel of Fig \ref{fig:xxtau}. The broad central peak is at $f_1$ = 6.38(01) c/d (equivalent to 3.76 h), the second most significant at $f_2$ = 5.38(06) c/d (4.46 h), the third at $f_3$ = 7.36(01) c/d (3.26 h) and others at $f_4$ = 4.31(06) c/d (5.57 h), $f_5$ = 8.29(06) c/d (2.90 h) and $f_6$ = 9.29(01) c/d (2.58 h). We noted that $f_3$ agrees well with the period favoured by \citet{rodriguez-gil2005}. However, the periodogram presented in that article (their fig. 14) shows a number of similarly strong aliases that are not properly discussed by the authors. A comparison with the periodogram for our data shows that all our significant frequencies coincide with the peaks in their periodogram. Thus, we find that, from the periodograms, we have six valid frequencies. 
However, folding our data with each of the corresponding periods for frequencies $f_2$, $f_4$, $f_5$ and $f_6$ showed systematic deviations from the fit (e.g. in the sense that a data set from one specific night presented a systematic offset), while for $f_1$ and $f_3$ the distribution of all data was consistent with random noise.
As mentioned above, each broad peak in our periodogram is formed by a series of narrow peaks, and thus each of the broad peaks for $f_1$ and $f_3$ contains several valid frequencies, which are displayed in Table \ref{tab:xxtau}. We thus here give the respective strongest ones of those as fiducial frequencies, but have to keep in mind that more valid possibilities within $3\sigma \sim$ 0.2 c/d exist. Defining $T_{0}$ as the red-to-blue crossing time in the radial velocities sinusoidal fit, the ephemeris for $f_1$ then is 
\begin{equation}
{\rmn{HJD}}=2\,458\,484.620(23) + 0^{{\rmn{d}}}.0.15664(28)~{\rmn{E}},
 \label{eq:xxtau1}
 \end{equation}
\noindent
and for $f_3$ is 
 \begin{equation}
  {\rmn{HJD}}=2\,458\,484.632(45) + 0^{{\rmn{d}}}.13588(21)~{\rmn{E}}.
 \label{eq:xxtau2}
 \end{equation}

\begin{table}
\centering

%\resizebox{0.5\textwidth}{!}{\begin{minipage}{\textwidth}
 \caption{Values of the possible orbital frequencies and its respective orbital period for XX Tau within one FWHM of the broad peaks centered at $f_1$ = 6.38 c/d and $f_3$ = 7.36 c/d in the periodogram shown in Fig. \ref{fig:xxtau} (d).}
 % \scalebox{0.78} {
 
\begin{tabular}{@{}crrr}
\hline
 \multicolumn{1}{c}{$f$} &
 \multicolumn{1}{c}{{$\rmn{P}$}} &
   \multicolumn{1}{c}{$f$} &
  \multicolumn{1}{c}{$\rmn{P}$}\\
   \multicolumn{1}{c}{c/d } &
  \multicolumn{1}{c}{ days } &
 \multicolumn{1}{c}{c/d} &
  \multicolumn{1}{c}{days} \\
\hline
\multicolumn{1}{c}{$f_1$}&        & 6.48(02) & 0.15432(45)     \\  
6.14(01) & 0.16289(30)            & 6.51(02) & 0.15364(45)     \\
6.16(01) & 0.16223(30)            & \multicolumn{1}{c|}{$f_3$}&\\
6.19(01) & 0.16158(30)            & 7.19(01) &  0.13908(22)    \\  
6.21(01) & 0.16093(29)            & 7.21(01) &  0.13863(22)    \\  
6.24(01) & 0.16033(29)            & 7.24(01) &  0.13816(22)    \\  
6.26(01) & 0.15971(29)            & 7.26(01) &  0.13768(22)    \\  
6.29(01) & 0.15904(29)            & 7.29(01) &  0.13724(21)    \\  
6.31(01) & 0.15848(29)            & 7.31(01) &  0.13678(21)    \\  
6.34(01) & 0.15782(28)            & 7.34(01) &  0.13632(21)    \\  
6.36(01) & 0.15724(28)            & 7.38(01) &  0.13544(21)    \\  
6.41(01) & 0.15609(28)            & 7.41(01) &  0.13499(21)    \\  
6.43(01) & 0.15546(28)            & 7.43(01) &  0.13452(21)    \\  
6.46(01) & 0.15485(27)            & 7.46(01) &  0.13408(20)    \\

\hline
                        
\end{tabular}
%}
\label{tab:xxtau}
\end{table}
 
\noindent
As example, in Fig. \ref{fig:xxtau} we show the light curve and the radial velocities folded with the period from Eq.\ref{eq:xxtau2}, since this is the value favoured by \citet{rodriguez-gil2005}. The sinusoidal fit corresponding to this period exposes a wide semi-amplitude $K$ = 167(12) km/s and the systemic velocity $\gamma$ = -22(10) km/s is slightly blueshifted. In the case for $f_1$ the fit parameters are $K$ = 160(16) km/s and $\gamma$ = -10(8) km/s. The photometric data (Fig. \ref{fig:xxtau} top) does not show any modulation for either period.

One possibility for the modulation found \citep{rodriguez-gil2005} not being present in our photometric data is that strong flickering on larger time-scales than in the LCO data mimicked a periodic signal in their data. However, this flickering would then to have maintained these same properties over a time span of six nights, which appears unlikely. Furthermore, the proximity to the spectroscopic signal is suspicious. A different possibility is that the system was caught in two different brightness states. In that case, the data from \citet{rodriguez-gil2005} could correspond to a state with a fainter accretion disc, where the bright spot would be more dominant and thus could produce an orbital hump in the light curve. In brighter, optically thick, accretion discs, on the other hand, the bright spot is typically much diminished or even not visible at all \citep{warner2003book}. Still, the long-term light curve
from \citet{lifevii}, if noisy, is consistent with a constant brightness over a range of about 1.5 yr. However, comparing our spectroscopic data with that of \citet{linda2005xxtau}, we find that the equivalent width of the H$\alpha$ line in our data with 28 {\AA} amounts to only roughly half the value that they found in their data (52 {\AA}). This points to a difference in the disc brightness, with a stronger line indicating a fainter disc. 
Unfortunately, we do not have any calibrated photometric information for either the \citet{linda2005xxtau} nor the \citet{rodriguez-gil2005} data. However, reviewing above evidence and sorting the dates, we find that XX Tau likely inhabited a fainter disc in late October and early November 2002 \citep{rodriguez-gil2005}), and in
January 2003 \citep{linda2005xxtau}, but a brighter
disc in December 2013 and January 2014 (our photometric data) and in December 2018 and January 2019 (the spectroscopic data). The long-term CTIO data covers the range
from November 2013 to April 2015. This timeline is thus
consistent with the possibility that XX Tau at some point
between January 2003 and November 2013 (at least once)
underwent a change from a low mass-transfer state with a
faint disc to a higher mass-transfer state with a brighter one.

%%%%%%%%%%%%%%%%%%%%%%%%%%%%%%%%%%%%%%%%%%%%%%%%%%%%%%%%%%%%%%%%%%%%%%%%

%%%%%%%%%%%%%%%%%%%%%%%%%%%%%%%%%%%%%%%%%%%%%%%%%%%%%%%%%%%%%%%%%%%%

\subsection{CQ Vel (1940)}
This nova reached its maximum brightness, $m_\mathrm{pg}$ = $\rmn{9^{m}}$ in April 19, 1940, being discovered on Franklin-Adams plates by C. J. Van Houten \citep{houten50cqvel}. It was categorized as a moderately fast nova with $t_\mathrm{3}$ = 53 d \citep{duerbeck81t3} and a large amplitude ($\rmn{A}_{v}$ $>$ $\rmn{13.1^{m}}$).
\noindent
The nova was recovered by \citet{woudt2001cqvel}, who performed high speed photometry in the field of a candidate for the nova proposed by \citet{duerbeck87Catalogue}. A strong flickering activity in a single, 4.07 h long, light curve was detected in an object 9 arcsec from the suspected position. Spectroscopic observations made by \citet{linda2005xxtau} using those coordinates confirmed the post-nova. They reported an equivalent width of H$\alpha$ line as 18 {\AA}, while from our new EFOSC2 spectra the value is 14.5 {\AA}.

\begin{figure}
% \centering  % this centres figure in column
  \resizebox{0.5\textwidth}{!}{\includegraphics[trim={12mm 0mm 0mm 0mm}, clip]{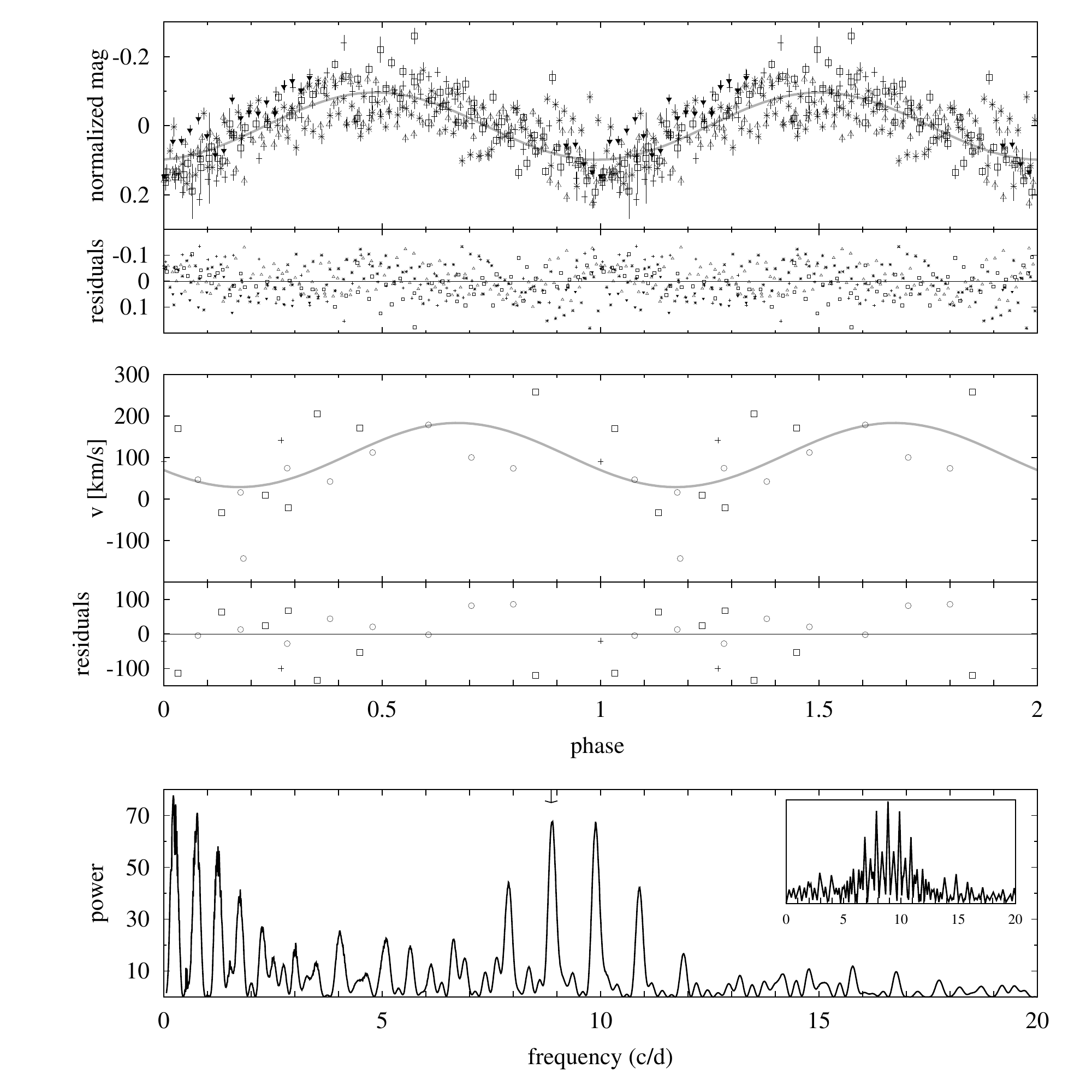}}
\caption{Top: Phased light curve for CQ Vel according to ephemeris (\ref{eq:cqvel1}), the sinus fit to the orbital modulation (grey line) and the residuals of the sine fit are shown. Middle: Radial velocity, the sinus fit and the residuals. Bottom: Periodogram of the photometric data including its spectral window centered to frequency pointed with the arrow.}
\label{fig:cqvel}
\end{figure}

Our light curves (Fig. \ref{fig:cqvel-lc}) show strong flickering activity as was seen by \citet{woudt2001cqvel}. The periodogram of the photometry (Fig. \ref{fig:cqvel}) shows two dominant frequencies at {\it f}$_{1} = 8.87$ and {\it f}$_{2} = 9.86$ c/d. Using both frequencies we found the following ephemerides for the photometric minima
 \begin{equation}
 {\rmn{HJD(min)}}=2\,456\,655.786(03) + 0^{{\rmn{d}}}.11272(12)~{\rmn{E}},
\label{eq:cqvel1}
\end{equation}
\noindent
for {\it f}$_{1}$ and the alternative: 
 \begin{equation}
 {\rmn{HJD(min)}}=2\,456\,655.782(06) + 0^{{\rmn{d}}}.0.10138(24)~{\rmn{E}}.
\label{eq:cqvel2}
\end{equation}
\noindent 
\textbf{The RMS scatter of the observed minima around ephemeris (\ref{eq:cqvel1}) is 0.0039 d and for (\ref{eq:cqvel2}) is 0.0088 d.} Both the radial velocity and the photometric data were folded with these ephemeris, however no significant differences were found. In the same way, the average semi amplitude ({\it K}) for the photometric phased data are practically identical within the errors, {\it K} = 0.0966(32) \textbf{mag} and {\it K} = 0.0932(34) \textbf{mag} respectively. 

We also note that, while a photometric sinusoidal signal could in principle be explained as the result of ellipsoidal modulation with orbital period twice the observed period, the radial velocities from our EFOSC2 spectra rule this out. The sine fit according to Eq. (\ref{eq:cqvel1}) yields a systemic velocity $\gamma$ = -106(15) km/s and a semi-amplitude {\it K} = 77(20) km/s. The radial velocities are displayed in Table \ref{tab:rv}.
 %One interesting thing is that the He I 6680 $\AA$ practically was missing ten years ago (see Fig. \ref{fig:spectra}).
 
 Under those circumstances, a decision which of the alternatives is the correct one must await more data. In any case, both of these periods place CQ Vel within the period gap.

%%%%%%%%%%%%%%%%tabla de efemerides

\begin{table}
\centering

%\resizebox{0.5\textwidth}{!}{\begin{minipage}{\textwidth}
 \caption{Epochs for the eclipsing systems.}
 % \scalebox{0.78} {
 
\begin{tabular}{@{}crrr}
\hline
 \multicolumn{1}{c}{Object} &
 \multicolumn{1}{c}{{$\rmn{E}$}} &
   \multicolumn{1}{c}{$\mathrm{HJD}$} &
  \multicolumn{1}{c}{$\mathrm{O-C}$}\\
   \multicolumn{1}{c}{ } &
  \multicolumn{1}{c}{ } &
 \multicolumn{1}{c}{$-2\,450\,000$ d} &
  \multicolumn{1}{c}{d} \\
\hline
X Cir      & $-$3041 & 6696.7977 & 0.0044 \\
           &$-$3028 & 6698.8051 & 0.0038 \\
           &$-$2931 & 6713.7858 & 0.0020 \\
           &$-$2879 & 6721.8137 & $-$0.0020 \\
           &$-$2782 & 6736.7933 & $-$0.0050 \\
           &$-$2058 & 6848.6214 & $-$0.0056 \\
           &$-$26   & 7162.4907 & 0.0020 \\
           &$-$25   & 7162.6453 & 0.0021 \\
           &$-$24   & 7162.8000 & 0.0023 \\
           &$-$19   & 7163.5690 & $-$0.0010 \\
           &$-$12   & 7164.6497 & $-$0.0015 \\
           &0       & 7166.5036 & $-$0.0011 \\
           &311     & 7214.5411 & $-$0.0005 \\
DY Pup     & 0 & 6658.6478 & 0.0001 \\
           & 1 & 6658.7873 & -0.0001\\
           & 7 & 6659.6245 & 0.0001 \\
 
\hline
                        
\end{tabular}
%}
\label{tab:t2}
\end{table}

\subsection{RW UMi (1956)}
\noindent
In the \citet{cvCatalogue2003} catalogue, RW UMi is listed as the nova with the shortest orbital period that is not marked as ``uncertain". The value of $P = 0.05912(15)$ d is based on photometric data taken in 14 nights over a total time range spanning almost four months \citep{retter2001_rwumi}. The period corresponds to a sinusoidal variation in the light curve with an amplitude of $\rmn{0.05^{m}}$ in white light. Later photometric studies by \citet{bianchini2003-RWUmi}
and \citet{tamburini-2007RWumi} found a number of other periodicities with larger amplitudes, suggesting that RW UMi is an intermediate polar showing quasi-periodic oscillations. They also found that the brightness of the nova is still declining at an approximate rate of 0.03 mag/yr as measured from the year 1988 to 2006. The existence of multiple photometric periods lets the identification of the reported value with an orbital modulation appear ambiguous, thus motivating the present spectroscopic study.

\begin{figure}
% \centering  % this centres figure in column
  \resizebox{0.5\textwidth}{!}{\includegraphics[trim={11mm 38mm 0mm 0mm}, clip]{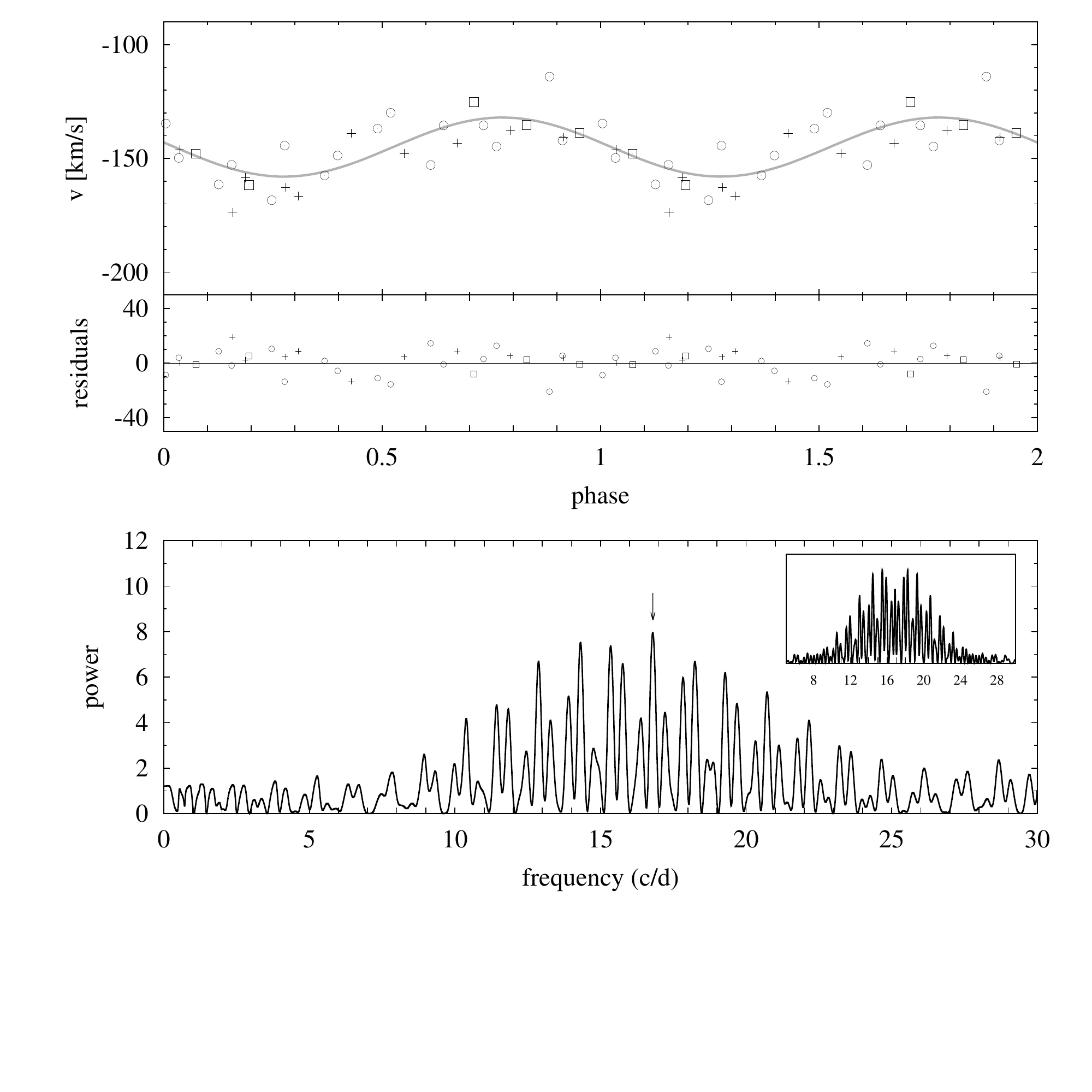}}
\caption{Top: Phase-folded radial velocities of RW UMi, the corresponding sine fit according to Eq.~(\ref{eq:rwumieph}) and the residuals. Different symbols indicate data from different nights. Bottom: Scargle periodogram of the radial velocity data.
The arrow marks the highest peak at {\it f} = 16.80(10) c/d. As inset plot is shown the spectral window centered at this frequency.}
\label{fig:rwumi} 
\end{figure}

Compared to other post-novae, the emission lines in RW UMi are relatively strong, with H$\alpha$ presenting an equivalent width of $-$17 {\AA}. However, the line profile is complex and non-Gaussian, with a broad base and a more narrow main component, with likely more than one source contributing to the
latter, as evidenced by its markedly variable shape 
(Fig.~\ref{fig:spectra}). Additionally, we were unfortunate in that the longest data set counted with the worst weather conditions of the three nights, resulting in significantly diminished S/N. Finally, obtaining a conclusive radial velocity curve is further complicated by the line presenting a comparatively small Doppler shift. In view of these difficulties, we employed a number of methods to determine the radial velocities, measuring different parts of the line or using the manual cross-correlation mentioned above. However, we found that in the end the clearest curve was produced by
fitting a single Gaussian function to the full line profile. The corresponding Scargle periodogram is presented in Fig.~\ref{fig:rwumi}. The strongest peak corresponds to a frequency {\it f} = 16.80(10) c/d, with the uncertainty being estimated by assuming a normal distribution. This translates to
a period $P = 0.0595(4)$ d, which, within one sigma, is identical to the photometric period of \citet{retter2001_rwumi}. Our periodogram shows several aliases that are close in strength to the main peak, and, taken on its own, it does not represent sufficient evidence to assign the orbital period. However, the good agreement with the photometric period strongly suggests that this indeed reflects the orbital motion of the system.

A sine fit to the radial velocity data according to Eq.~(\ref{eq:fit-rv}) yields the aforementioned small semi-amplitude $K = 13(1)$ km/s and a markedly blueshifted systemic velocity $\gamma = -145(1)$ km/s (lower plot in Fig.~\ref{fig:rwumi}). Choosing the red-to-blue crossing of the radial velocities as the zero point of the phase-folded curve and using the more precise photometric period yields a formal ephemeris of
\begin{equation}
\mathrm{HJD} = 2\,457\,196.4397(10) + 0^\mathrm{d}.059\,12(15)~{\rmn{E}}~,
\label{eq:rwumieph}
\end{equation}
although, considering the complex nature of the line profile, it is unlikely to correspond to the superior conjunction of the white dwarf.

%%%%%%%%%%%%%%%%tabla de efemerides

\begin{table}
\centering
%\resizebox{0.5\textwidth}{!}{\begin{minipage}{\textwidth}
%\begin{minipage}{1\columnwidth}
 \caption{Ephemerides for the eclipsing systems whose orbital periods could be confirmed by the CTIO observations \citep{lifevii}. $\mathrm{T_{0}}$ refers to minima of eclipses.
}
  \scalebox{1.1} {
\begin{tabular}{@{}lrr}
\hline
 \multicolumn{1}{c}{Name} &
  \multicolumn{1}{c}{$T_{0}$} &
  \multicolumn{1}{c}{$P_\mathrm{{orb}}$} \\
  \multicolumn{1}{c}{} &
  \multicolumn{1}{c}{$\mathrm{HJD-2\,400\,000}$} &
  \multicolumn{1}{c}{d} \\
\hline                 
OY Ara    &  56516.5722(10)&  0.155390(30)   \\
V849 Oph  &  48799.7412(18) &  0.17275611(06)  \\
WY Sge    &  47059.8678(04)  & 0.153634547(10) \\
V728 Sco  &  56015.8066(09) & 0.13833866(18)  \\

\hline
                        
\end{tabular}
}

%\end{minipage}
\label{tab:improvePorbs}
\end{table}

%%%%%%%%%%%%%%%%%%%%%%%%%%%%%%%%%%%%%%%%%%%%%%
\subsection{Improved ephemerides of eclipsing novae with previously known orbital periods}
\label{subsec:improvePorb} 
%%%%%%%%%%%%%%%%%%%%%%%%%%%%%%%%%%%%%%%%%%%%%%
We present refined orbital periods of WY Sge, V728 Sco, OY Ara and V849 Oph from CTIO data, which, by chance, have occasionally been caught during eclipse phases, showing fainter brightness than normally. The resulting ephemerides are listed in Table \ref{tab:improvePorbs}. Their epochs and O-C values, together with previously available literature and their references are listed in Table \ref{tab:Ectio}. Because the CTIO data consist of only two data points in any given night, they do not necessarily correspond to the central part of and eclipse, and thus the corresponding O-C deviations are larger in average than those from published photometry. Despite this, due to the larger time intervals covered now the new periods are more accurate.

We also have performed searches for periodicity in the other novae included in the CTIO data, V500 Aql, HS Pup, V1059 Sgr and V373 Sct, without finding any significant photometric periodicity.

%%%%%%%%%%%%%%%%%%%%%%%%%%%%%%%%%%%%%%%%%%%
\section{The Orbital period distribution of novae}
\label{sec:Porb-distribution}
%total: 92, al 2013: 78, se descartaron 5, pero esta lista no contenia a rs car
%nuevos nuestros: 6, nuevos desde 2013 de otros:12 . nuestra muestra equivale al 50% de incremento de la muestra desde 2013
\noindent
The current sample analysed here contains 92 orbital periods. From the sample listed by \citet[][]{tappertIII}, we selected those periods that satisfied the criteria defined below, giving a total of 74 periods, to which six new periods presented here were added, together with those new periods listed by \citealt{cvCatalogue2003} (version 7.24, 2016) since 2013. 
  
Here we present an analysis of the observed orbital period distribution of novae and compare it to simulated distributions both from the literature and with a newly established one that takes into account consequential angular momentum loss.

\subsection{Observed period distribution}
\label{subsec:obs-distribution}
\begin{figure}
\centerline{\includegraphics[width=.48\textwidth]{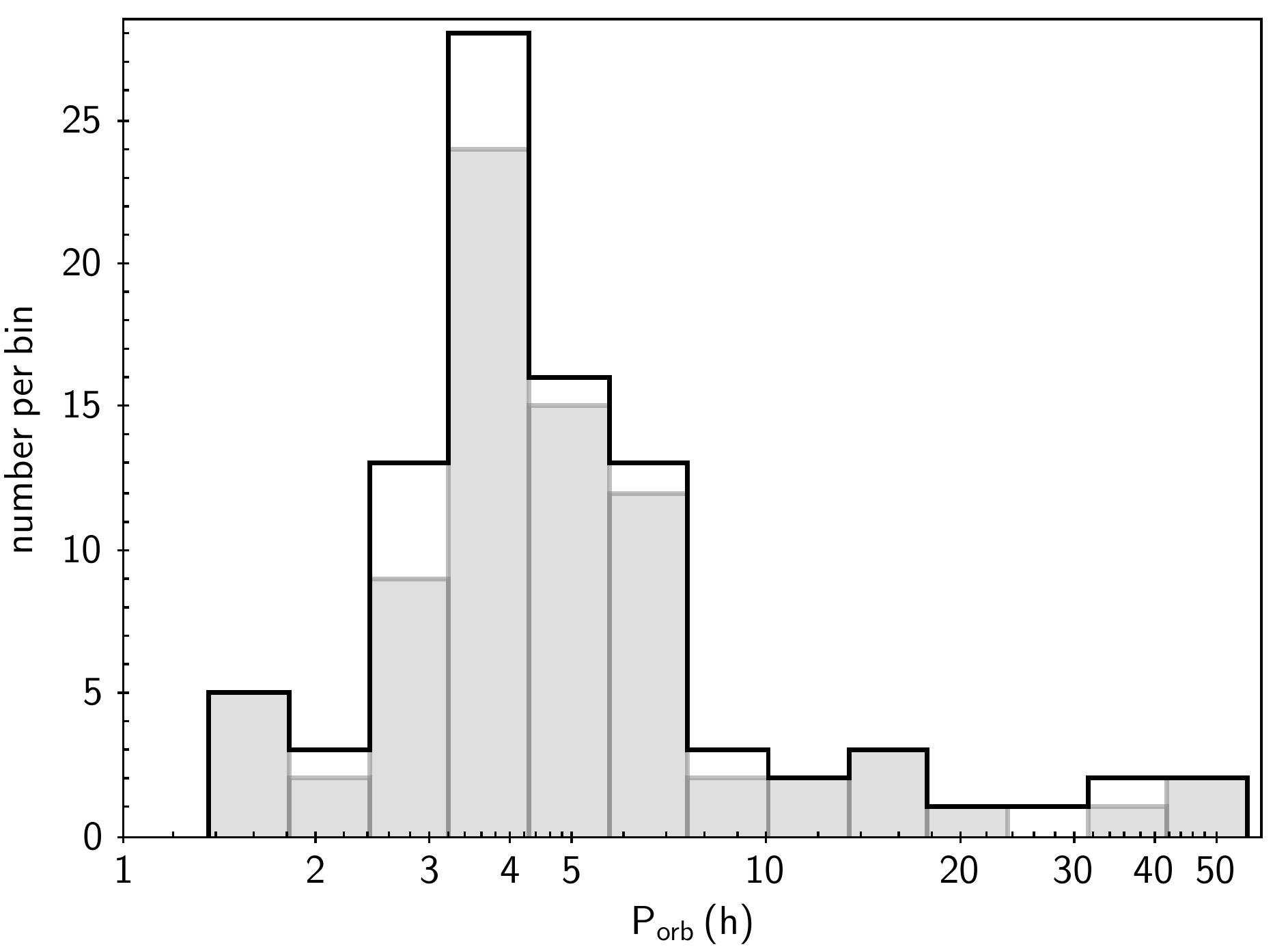}}
%\centerline{\includegraphics[width=.48\textwidth]{cumulative_porb.pdf}}
\caption{The current orbital period distribution of the novae on logarithmic scale (solid black line) in comparison with the distribution published previously by \citet[][ grey blocks]{tappertIII}.}
%Bottom: Cumulative distribution for both samples (black: this work, dashed: previous sample). }
\label{fig:histograma}
\end{figure}

%%%%%%%%%%%%%%%%%%%%%%%%%%%%%%%%%%%%%%%%%
% la tabla final esta en cataligos/cne+p3_claus2013+newsP_txt_sort2_CIERTOS.dat quitando el dato de v1017 sgr con P=137 d y considerando a cq vel.
\noindent
We used the catalogue of \citealt{RK2003} (version 7.24, 2016) to gather the period information on the novae included here. We excluded objects from the sample if their tabulated periods: (a) 
were not sufficiently coherent and might be attributed to 
QPOs; (b) might be caused by ellipsoidal variations at twice the orbital period; and (c) were based on data that has never been published. In addition, we (d) excluded objects for which the CV classification is not confirmed, with the data allowing for alternatives (e.g., in the case of light curves showing comparatively smooth sinusoidal variations that could also originate in pulsating stars). Table \ref{tab:uncertainP} presents the 24 novae that were excluded from the sample, based on above criteria. To the such established distribution we added our own results presented in the previous section. We have also included the novae RS Car, IL Nor, V2572 Sgr, XX Tau and CQ Vel, in spite of the fact that in those systems we cannot distinguish between more than one possible values for the orbital period. However, the periods are sufficiently close to correspond to the same period bin in the histogram, so that the overall distribution is identical for either of the alternatives. These novae are marked as ``provisionals'' in the Table used (\ref{tab:Porb}) for resulting distribution presented in Fig. \ref{fig:histograma}.  

% la tabla final esta en cataligos/cne+p3_claus2013+newsP_txt_sort2_CIERTOS.dat quitando el dato de v1017 sgr con P=137 d y considerando a cq vel% Given these points, seven novae were rejected from the \citeauthor{tappertIII}'s sample (DI Lac, GI Mon, V842 Cen, V382 Vel, V2362 Cyg and RW Umi), which our observed sample is compared.
 %de la muestra que considero claus, en esa distribucion no fueron considerados DI Lac (c), GI MOn (c), V842 Cen, V382 Vel, V2362 Cyg y RW Umi. 
 
Comparing the current distribution with the one published by \citealt{tappertIII} (in Fig. \ref{fig:histograma} are shown as a solid black line and grey blocks respectively) and using their same criteria to analyse the sample , i.e., considering the period gap as the range between 2.15 to 3.18~h \citep{knigge2006periodgap}, it is evident that both follow the same trend, with a strong maximum in the range of  3 -- 4 hr. In the new distribution most of the periods are above the period gap, corresponding to 79 per cent (equivalent to 72 objects), out of which 45 systems have $P_\mathrm{{orb}}$ > 4 h, equivalent to $\sim50\%$ of the total sample. The peak in the 3 -- 4 hr period range becomes more pronounced, concentrating 34 percent of the total sample (equivalent to 31 novae). On the other hand, eight per cent of the post-novae, corresponding to seven systems, are below the period gap and 14 per cent are in the period gap (corresponding to 13 systems).

 %%%%%%%%%%

\begin{table}
%\centering

%\resizebox{0.5\textwidth}{!}
\begin{minipage}{1\columnwidth}
 \caption{Novae with uncertain published $P_\mathrm{{orb}}$. The last column indicates the exclusion criterion as defined in the text.}
 % \scalebox{1} {
 
\begin{tabular}{@{}llcll}

\hline
  \multicolumn{1}{|l|}{Name} &
  \multicolumn{1}{l|}{$P_\mathrm{{orb}}$ (hr)} &
  \multicolumn{1}{c|}{Outburst}  &
  \multicolumn{1}{l|}{Ref.} &
  \multicolumn{1}{l|}{Cause}\\
\hline 
V705 Cas          & 5.47 & 1993    & (1) & (c)\\
V842 Cen         &3.94  & 1986    &  (2)  & (a) \\   
V2274 Cyg         & 7.2 & 2001     &  (3)  & (c) \\
V2362 Cyg         & 1.58 & 2006    &  (4) & (c)\\
V2491 Cyg         & 17  & 2008     &  (5) & (a)\\
DM Gem            & 2.95  & 1903   &  (7) & (a)  \\
DI Lac            & 13.05 & 1910   & (8) & (c)\\
DK Lac            & 3.11 & 1950    & (9) & (a)\\  
U Leo             & 3.21 & 1855    & (10) & (d)    \\ 
GI Mon            & >4.8 & 1918    & (6), (7) & (a)\\
LZ Mus            & 4.06 & 1998    & (11) & (c)  \\
V400 Per          & 3.84 & 1974    & (7) & (a) \\
%  KT Eri           & 700d & 2009    &   \\
%HZ Pup            & 5.11 & 1963    &  &    \\
V445 Pup          & 15.62 & 2000   & (12) & (a) \\
V574 Pup         & 1.13 & 2004    & (13) & (b)     \\
V1186 Sco         & 1.39 & 2004    & (3)   & (c) \\
V1324 Sco         & 3.8 & 2012     & (14) & (c)\\
V726 Sgr          & 19.75 & 1936   & (15) & (d)  \\
V999 Sgr         & 3.64 & 1910    & (15) & (b)    \\
V1174 Sgr         & 7.42 & 1952    & (15) & (d)   \\
V4077 Sgr          & 3.84 & 1982    & (16) &(c)\\
V5582 Sgr        & 3.76 & 2009    & (15) &  (b) \\
V5980 Sgr          & 30.34 & 2010   & (15) & (b)\\
V382 Vel           &3.79  & 1999    & (17), (18), (19) & (a)\\
PW Vul            & 5.13 & 1984    & (20) & (c)\\

\hline
\end{tabular}
%}
\\
References:\\
 (1) \citealt{v705casretter95telegram},  (2) \citealt{woudt2009V842cen}, (3) \citealt{RK2003} (4) \citealt{v2362cygtelegram}, (5) \citealt{zemko2018} (6) \citealt{woudt2004}, (7) \citealt{rodriguez-gil2005}, (8) \citealt{goransky97}, (9) \citealt{dklac2007},  \citealt{dklacnoP2011}, (10) \citealt{downes89}, (11) \citealt{lzmus}, (12) \citealt{v445pup2010}, (13) \citealt{walter2012}, (14) \citealt{v1324sco}, (15) \citealt{mroz2015}, (16) \citealt{diaz97}, (17) \citealt{woudt2005}, (18) \citealt{balmanV383vel}, (19) \citealt{eganv382vel}, (20) \citealt{pwvul87} \\

\end{minipage}
\label{tab:uncertainP}
\end{table}                                                                                                                                        

%%%%%%%%%%%%%%%%%%%%%%%%%%%%%%%%%%%%%%%%%%%
\subsection{Simulation}
\label{subsec:simulations}
%%%%%%%%%%%%%%%%%%%%%%%%%%%%%%%%%%%%%%%%%
\noindent
We generated an initial main-sequence plus main-sequence (MS+MS) binary population of $10^{9}$ systems with the following assumptions: initial-mass function of \citet{kroupaetal93-1} for the mass of the primary star;  flat initial mass-ratio distribution for the mass of the secondary star \citep{sanaetal09-1}; distribution of initial orbital separations ($a$) flat in $\log a$ ranging from $a = 3$ to $10^{4}\Rsun$ \citep{popovaetal82-1,kouwenhovenetal09-1}; constant star formation rate within the age of the Galaxy ($13.5\times10^{9}$\,yr, \citealt{pasquinietal04-1}); solar metallicity; and no eccentricity.

The binary-star evolution code (BSE) from \citet{hurleyetal02-1} was used to evolve the systems until the end of the common-envelope phase, i.e. until the close but detached WD+MS binaries (which are the direct progenitors of CVs) are formed. A common-envelope efficiency of $\alpha_{\mathrm{CE}}=0.25$ was assumed \citep{zorotovicetal10-1} and the binding energy parameter $\lambda$ was computed assuming that the recombination energy stored in the envelope does not contribute to the ejection process \citep{zorotovicetal14-1}. After this phase, the WD+MS systems were evolved using the CV evolution code developed by us and described in \citet{schreiberetal16-1} and \citet{zorotovicetal16-1}. It is based on the disrupted magnetic braking model for systemic angular momentum loss (AML), i.e., AML that is present even in the absence of mass transfer, due to gravitational radiation and magnetic wind braking (the latter only for CVs above the period gap). Inflation of the radius of the secondary star as a consequence of mass transfer is incorporated by using the observed mass-radius relation and the scaling factors for systemic AML from \citet{kniggeetal11-1}.
This code also takes into account the consequential AML produced by mass transfer and nova eruptions after the CV phase begins. Two models for consequential AML due to nova eruptions were included: the classical non-conservative model from \citet{king+kolb95-1} and the empirical model from \citet{schreiberetal16-1}. The latter predicts a smaller number of CVs, mainly because systems with low-mass WDs are driven into a dynamically unstable mass transfer regime and merge. As shown in \citet{schreiberetal16-1}, this has an effect not only on the WD mass distribution but also on the distribution of orbital periods. Here we want to test if there is also an effect on the predicted orbital period distribution of post-nova systems. 

Once the simulated populations of CVs have been generated, the probability of observing a nova eruption was computed for each system. This probability is inversely proportional to the nova recurrence time $P_{\mathrm{rec}}$, which can be written as:
\begin{equation}
P_{\mathrm{rec}} = m_{\mathrm{acc}}/\dot{M},
\end{equation}
where $\dot{M}$ is the mass transfer rate and $m_{\mathrm{acc}}$ is the accreted mass needed to produce a nova outburst. For each system we derived the value for $m_{\mathrm{acc}}$, which depends on the WD mass, the mass transfer rate, and the core temperature, based on \citet[][ interpolating their table 2]{yaronetal05-1}, who presented models for different fixed core temperatures. \citet{townley+bildsten04-1} found that the equilibrium core temperatures of WDs are below $10^{7}$\,K in typical CVs, and \citet{chenetal16-1} compared the observational data of novae in the M31 galaxy with the models from \citet{yaronetal05-1} and preferred the low temperature models. We have therefore chosen the values of $m_{\mathrm{acc}}$ listed by \citet{yaronetal05-1} for their models with the minimum core temperature ($10^{7}$\,K).

We also defined systems that experience more than a nova eruption in a century as recurrent novae \citep[e.g.][]{sharaetal18-1}. This means that if the computed recurrence period of a system in our simulation is less than 100 years, more than one nova eruption could be observed during that period of time. Therefore, we set an upper limit for the detection probability $\mathcal{P}_\mathrm{det} = (P_{\mathrm{rec}}[\mathrm{yr}])^{-1}$ of 0.01, which corresponds to a recurrence period of 100 years, to avoid counting recurrent novae more than once in the simulated period distribution. 

Systems in which the mass of the donor star falls below $0.05\Msun$ were eliminated from our simulated sample, because their mass-radius relation is not well constrained \citep[e.g.][]{kniggeetal11-1}. This has virtually no effect on the simulated distribution of orbital periods, because CVs with low-mass donors (below the brown-dwarf mass limit) have very low mass transfer rates which translate into very long recurrence periods, i.e. extremely low probabilities of being detected as post-nova systems. 
We also excluded CVs that experienced a thermal time-scale mass transfer phase, i.e. systems with initially massive donors ($M_2 \gappr 1.5 \Msun$), because the evolution during this phase is not well understood \citep[e.g.,][]{nomotoetal79-1,hachisuetal96-1}, and it is especially not clear how the mass of the WD could be affected. However, these systems should make up a small percentage of the current CV population \citep[$\sim5\%$;][]{pala2020}.

\subsection{Comparison}

\label{subsec:comparison}
\begin{figure}
% \resizebox{0.5\textwidth}{!}{\includegraphics[trim={left botton right up}, clip]{histo.pdf}}
  \resizebox{0.5\textwidth}{!}{\includegraphics[trim={23mm 52mm 10mm 33mm}, clip]{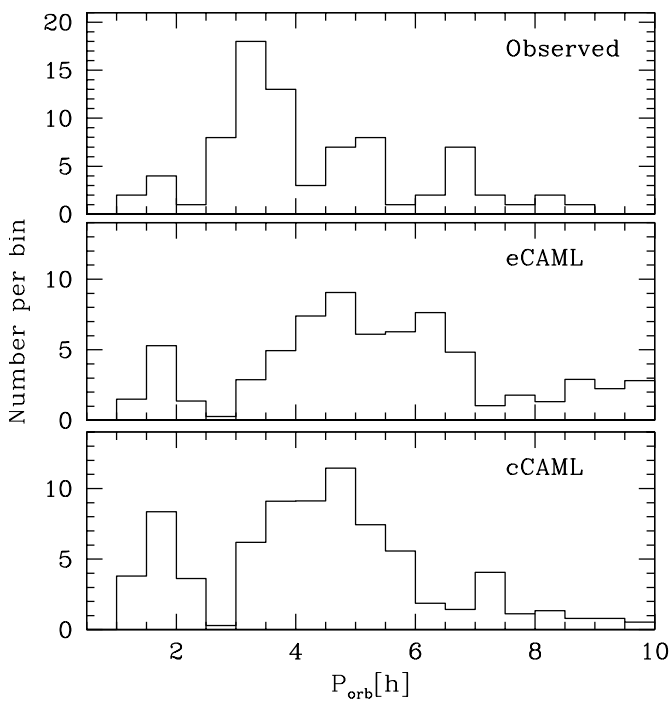}}
\caption{Orbital period distribution of post-nova systems. From top to bottom: observed systems (this work), simulation assuming the empirical consequential AML model from \citet{schreiberetal16-1} and simulation assuming the classical consequential AML model from \citet{king+kolb95-1}.}
\label{fsim}
\end{figure}

The predicted orbital period distributions were constructed using the same bins as for the observed distribution, adding the detection probabilities for all the simulated systems within that period range, and normalizing to the observed number of systems. The results are shown in Fig.\,\ref{fsim}. 
The two models of AML predict broadly similar distributions, and both show the majority of systems above the gap, in keeping with the observed distribution (top panel).
%with orbital periods between $\sim 3$ and $8$ hours.
The classical non-conservative model from \citet[][ bottom panel]{king+kolb95-1} predicts that $\sim18\%$ of novae should be observed below the orbital period gap, $\sim2-3\%$ in the gap, and $\sim79-80\%$ above it. For the simulations that assume the empirical model from \citet[][ middle panel]{schreiberetal16-1}, the expected fractions are $\sim9-10\%$ below the gap, only $\sim1\%$ in, and $\sim89-90\%$ above. 
%IRMA   7% below 14% in 79% above 
%cCAML 18.04  2.50 79.46 
%eCAML  9.79  0.65 89.57 
%Although the classical model predicts much more CVs below the gap than the empirical model, the low mass transfer rates of these systems make their recurrence periods very long, which translates into a very low probability of observing a nova outburst. 
The empirical model is therefore in better agreement with the observations, regarding the fraction of systems that we expect to observe below the orbital period gap. However, this conclusion should be taken with caution because we are dealing with low-number statistics and our poor knowledge of CV evolution. 

The over-prediction of systems below the gap in the two simulations with respect to the observed distribution might be explained by poor constrained aspects of CV evolution. A key parameter that might affect the simulated distribution is the assumed core temperature. As explained before, we used a constant core-temperature for the calculations of the accreted mass needed to produce a nova outburst ($m_{\mathrm{acc}}$). As calculated by \citet{yaronetal05-1}, colder WDs should need to accrete more mass before triggering the eruption of a nova. Given that the evolution of a CV towards shorter periods is mainly driven by systemic AML, the lifetime of a system is much shorter above the gap than below it due to the efficiency of magnetic braking. This implies that CVs below the gap are, on average, older than CVs above the gap. Therefore, the core temperature of the WDs in CVs below the gap should be lower, on average, because they have had more time to cool. Also, it is not clear whether the accretion process can affect the temperature of the core of the WD \citep[e.g.,][]{cumming02,townley+bildsten04-1,townsley+gaensicke09-1}. If the core temperature can increase as a result of mass accretion, this increase should be larger for CVs above the gap, in which the accretion rate is higher. Combining these two effects implies that by assuming a constant WD core temperature for all the systems we are probably underestimating the value of $m_{\mathrm{acc}}$ needed to trigger a nova eruption for CVs below the gap, which means that their contribution to the predicted post-nova population is overestimated. A more accurate derivation of $m_{\mathrm{acc}}$ that depends on the core temperature for each WD is beyond the scope of this paper, but the effect of such an improvement on the models would probably be to reduce the fraction of novae predicted below the gap, for both models.

Another discrepancy with the observations is that our simulations predict an extremely low fraction of novae in the orbital period range that corresponds to the period gap, for both models of consequential AML (below $3\%$, while observationally it is $\sim14\%$). This is a direct consequence of assuming efficient magnetic braking for all CVs above the period gap. However, magnetic braking can become very inefficient for CVs containing WDs with strong magnetic fields. According to \citet{bellonietal19-1}, the WD magnetic field in strongly magnetized CVs can trap part of the wind from the donor reducing the loss of angular momentum through this wind. This implies that magnetic CVs above the gap have lower mass transfer rates than their non-magnetic counterparts, and have therefore less bloated donors. This translates into a shift of the upper edge of the gap towards shorter periods, or even a complete absence of the detached phase for CVs with the strongest WD magnetic fields. In other words, magnetic CVs can cross, or at least enter, the orbital period gap.
Indeed, the gap seems to be much less pronounced in the observed period distribution of magnetic CVs than in that of non-magnetic CVs \citep[e.g.,][their Fig.\,17]{ferrarioetal15}.
The fraction of magnetic WDs in CVs is known to be high (e.g., $\sim 33\pm7$ per cent in the first volume-limited sample of CVs, recently published by \citealt{pala2020}). Therefore, including a fraction of magnetic CVs in our simulation, with reduced magnetic braking model like the one described by \citet{bellonietal19-1}, could help reconcile the fraction of novae observed in the gap.

Regardless of the model, the main difference between our simulations and the observed period distribution is the presence of a peak in the number of observed systems with periods between 3 and 4 hours that our models do not reproduce. Above the period gap, the mass transfer rate depends mainly on the formalism assumed for magnetic braking. Here we assumed the \citet{rappaportetal83-1} prescription for $\gamma = 3$, with the normalization factor derived by \citet{kniggeetal11-1}. As can be seen in \citet[][ their figure\,2]{kniggeetal11-1}, this formulation predicts a reduction in AML when approaching the period gap from larger periods. The simulated mass transfer rates are therefore lower for systems in the period  range of $3-4\,$h compared to systems with larger periods, making their recurrence periods longer.  \citet{kniggeetal11-1} also showed that assuming a smaller value for $\gamma$ in the \citet{rappaportetal83-1} prescription for magnetic braking, or the formulation developed by \citet[][ which is the same as the \citet{andronovetal03-1} model in the unsaturated limit]{kawaler88-1}, would all predict an increase of AML towards shorter periods, which would transfer into larger mass transfer rates and smaller recurrence periods that could reconcile the predictions with the observations. 

The existence of a peak in the period distribution of post-nova system at $3-4\,$h, in addition to observational evidence of higher mass transfer rates for CVs in the same period range \citep{townsley+gaensicke09-1,palaetal17-1}, seems to indicate that the \citet{rappaportetal83-1} prescription with $\gamma = 3$ might not be the best approximation for magnetic braking in non-magnetic CVs. A similar conclusion is drawn in \citet{bellonietal19-1}, where the simulated mass transfer rates for non-magnetic CVs above the gap drastically disagree from observations (when assuming also $\gamma = 3$ in the \citet{rappaportetal83-1} prescription for magnetic braking), suggesting that AML caused by magnetic braking is not well understood. An interesting future work would be to derive the normalization factors, similar to what was done in \citet{kniggeetal11-1}, but for magnetic braking prescriptions that predict an increase of AML while approaching the gap from larger periods. This would allow us to test whether the peak in the observed period distribution can be reproduced by changing the formulation for magnetic braking only. 

%%%%%%%%%%%%%%%%%%%%%%%%%%%%%%%%%%%%%%%%%%%%%%%%%%%%%%
Comparing with the literature, the only theoretical orbital period distribution of novae previously published is that of \citet{townsleyBildsten2005}. They predict a strong peak in the range of $P_\mathrm{{orb}}$=3 -- 4 h,  but their cumulative distributions do not fit well for periods larger than 4 hours, which corresponds to $\sim50\%$ of the observed sample. As the same authors mentioned, they use a very simple CV population model, where the number of CVs at each period interval was taken from \citet{howelletal01} with a fixed WD mass, instead of evolving the systems from a binary population synthesis model. In order to obtain the mass transfer rate, they used the same prescription as we did for magnetic braking (i.e., \citealt{rappaportetal83-1} prescription with $\gamma = 3$), but with a different  mass-radius relation for the donor stars above the gap (also from \citealt{howelletal01}). The accreted mass needed to produce a nova outburst was based on \citet{townley+bildsten04-1} instead of \citet{yaronetal05-1}. It is therefore impossible to make a more detailed comparison between their models and ours, although the need to include magnetic CVs with reduced magnetic braking in order to reproduce the fraction of novae observed in the period gap is a common conclusion of both studies.
%%%%%%%%%%%%%%%%%%%%%%%%%%%%%%%%%%%%%%%%%%%%%%%%%%%%

\section{summary and conclusions} 
\label{sec:summary}
%%%%%%%%%%%%%%%%%%%%%%%%%%%%%%%%%%%%%%%%%%%%%%%%%%%%%
%%%%%%%%%%%%%%%%%%%%%%%%%%%%%%%%%%%%%%%%%%%%%%%%%%
\noindent
We have presented six new orbital periods and have reviewed and/or improved the periods for eight old novae, and discussed the resulting distribution of observed orbital periods with respect to theoretical predictions based on a binary population synthesis model. In the following we summarize the most noteworthy results and conclusions.

\renewcommand{\labelitemi}{$-$}
\begin{itemize}
\item %how many eclipsing systems
With X Cir, we report one new eclipsing nova, and with DY Pup, we confirm another one that was previously reported, but lacked the data to sustain such claim. Both have orbital periods in the  $3-4\,$h range, corresponding to the period regime that is dominated by high $\dot{M} $ objects. Comparing those systems, we find that the eclipses in DY Pup with a depth of $\sim\rmn{0.3^{m}}$ are considerably more shallow than those of X Cir that show an average depth of $\rmn{1^{m}}$, which could indicate that the latter object is seen at a somewhat higher inclination than the former. X Cir's high inclination could also possibly account for the spectral appearance that was interpreted by \citep{tappertIV} as a signature of a low $\dot{M}$ system \citep{warner86-inclinations}. 
\item For RS Car, IL Nor, V2572 Sgr, XX Tau and CQ Vel, there is still some ambiguity concerning the orbital period, with more than one possible values existing for both objects. Still, we can already conclude that CQ Vel, together with V363 Sgr are situated in the period gap, while IL Nor is placed below it, making it the oldest nova in that short-period regime. The detection of an orbital modulation in the light curve of V363 Sgr indicates that it is seen at a somewhat higher inclination than suspected by \citet{tappertIV}. 
\item For three targets (V2572 Sgr, CQ Vel and RW UMi) the orbital period was determined or confirmed by time-resolved radial velocity observations. For the confirmation of the orbital period for XX Tau we suggest trying by this technique, observing with a baseline larger than four hours.

\item In addition to short-term time-resolved photometric observations, we also used the CTIO data set, with a typical time resolution of 3 -- 4\, days, a by-product of a search for stunted dwarf nova-like outbursts in classical novae \citep{lifevii}. Our new period of V363 Sgr is entirely based on these data; they also enabled to derive a long-term orbital ephemeris of V2572 Sgr, and to improve the periods of other six novae (four with eclipses and two with orbital humps). 

%how do the new orbital periods distribute
\item We also present a statistic of all currently known orbital periods of novae, which are  distributed in the following way: 79 per cent are located above the gap, equivalent to 72 objects, $\sim50$ per cent of them ($=$ 45 objects) have $P_\mathrm{{orb}} > $ 4 h. Only seven systems are located below the period gap, corresponding to eight per cent of the sample, meanwhile 13 systems (14 per cent) were found within the period gap. It is worth mentioning here again that this distribution differs significantly from the one of all CVs, with the main differences being the low number of objects below the gap, the majority of the novae having period above the gap, and especially the peak located above the gap at 3 -- 4\,h that with the new data has become even more pronounced.

\item There are striking differences between the theoretically predicted period distribution of novae and the observed one. Population model calculations are in accordance with the observed number ratios of novae below, within and above the period gap, but they are not able to reproduce the rather narrow peak observed  at 3 -- 4\,h. Instead, they predict a more flat distribution in the range $3\,$h  $\leq P_\mathrm{{orb}}\leq$ 6 -- 8\, h. This implies that the prescription usually used for AML due to magnetic braking in CVs above the period gap might not be correct. 
%WE ARE TALKING ABOUT NOVAE NOT ABOUT CVS IN GENERAL. WE DO NOT SHOW THE DISTRIBUTIONS FOR CVS HERE. On the other hand, models predict much more CVs below the gap than observed, but due to the low mass transfer rates of these systems their nova recurrence periods are very long, which translates into a very low probability of observing nova eruptions in CVs below the period gap. 

\end{itemize} 
%Along the course of our ``Life after eruption'' series, we have achieved new determinations of the orbital periods for sixteen relatively faint old novae, which results in an increase of $\sim20$ per cent with respect to the total sample known in 2013, at the begin of the series.  60 per cent  of them refer radial velocity orbits and 40 per cent to  photometric variations, as orbital humps or eclipses. Only one of them turned out to be below the period gap, two are within it and thirteen are above the gap.
%% en resumen de proyectos de claus dice que se han encontrado 12 new Porb (aumento del~27%)..respecto a que muestra se hizo este %%? 

Finally, we would like to mention that a new generation of terrestrial survey telescopes will soon become into operation, for instance the Vera C. Rubin Observatory (previously referred to as the Large Synoptic Survey Telescope, LSST), which will observe a large portion of the entire sky every  $\sim3$ days, (a similar cadence as our CTIO set) revealing crucial information on the behavior of many not yet observed, or even not yet identified old novae. This way, we will finally obtain better statistics on the orbital period distribution and other unsolved questions addresses here, but still open.

\section*{ACKNOWLEDGEMENTS}
We are grateful to John Thorstensen for his detailed and valuable report. We thank Maja Vu\v ckovi\'c for suggestions related to the nova XX Tau. We give thanks to the European Southern Observatory (ESO) for the following observing runs in service mode: 087.D-0323(A), 088.D-0588(A) and 0102.D-0488(A). For RW UMi we give thanks to the Gran Telescopio Canarias (GTC) for the observing run GTC36-15A. 
IFM thanks to CONICYT-PFCHA/ Doctorado Nacional 2017-21171099 for doctoral fellowship. 
CT and NV acknowledge support form FONDECYT grant number 1170566. 
MZ acknowledges support from CONICYT PAI (Concurso Nacional de Inserci\'on en la Academia 2017, Folio 79170121) and CONICYT/FONDECYT (Programa de Iniciaci\'on, Folio 11170559).
MRS acknowledges financial support from FONDECYT grant number 1181404 and the N\'ucleo de Formaci\'on Planetaria (NPF). 

\section*{Data availability}
\noindent
The data underlying this article are available in the appendix section and the photometric data in its online supplementary material. 
%Radial velocity data used to derive the orbital periods for V2572 Sgr, XX Tau, RW UMi and CQ Vel are available in the appendix section (Table \ref{tab:rv}). The orbital periods used to compose the orbital period distribution analysed here are shown in Table \ref{tab:Porb}. The photometric data will be shared on reasonable request to the corresponding author.



\bibliographystyle{mnras}
\bibliography{ref} % if your bibtex file is called ref.bib


%%%%%%%%%%%%%%%%%%%%%%%%%%%%%%%%%%%%%%%%%%%%%%%%%%

%%%%%%%%%%%%%%%%% APPENDICES %%%%%%%%%%%%%%%%%%%%%

\appendix
\section{Extra material}
\label{sect:appendix}
\noindent
Individual light curves and spectra for the analysed novae are shown as extra material. The epochs for the eclipses and the radial velocity measurements are also presented in the follow tables.

%%%%%%%%%%%%%%%%%%%%%%%%%%%%%%%%%%%%%%%%%%%%%%%%%%
\begin{figure*}
% \centering  % this centres figure in column
  \resizebox{1\textwidth}{!}{\includegraphics[trim={0mm 00mm 0mm 0mm}, clip]{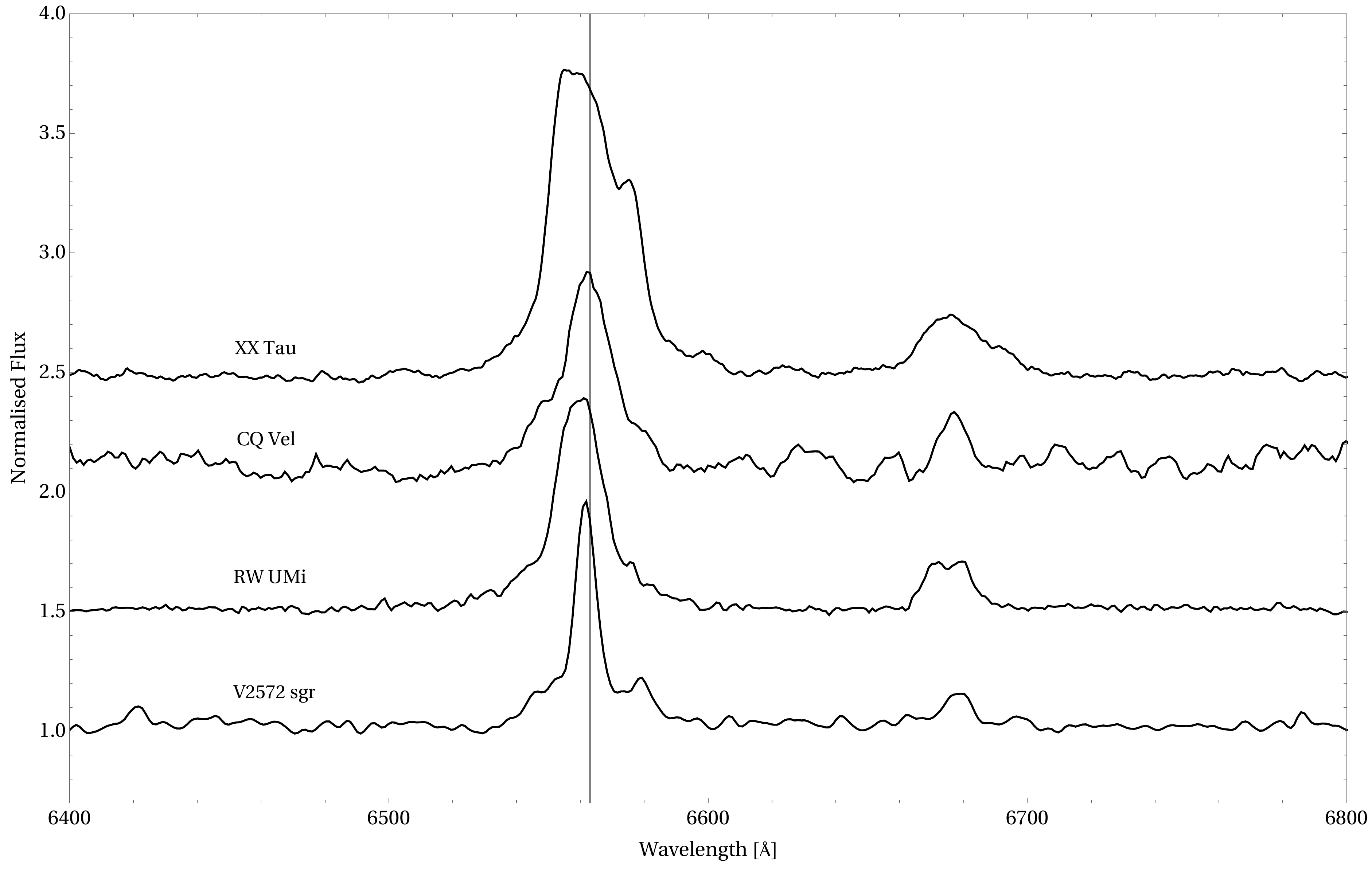}}
\caption{Normalised average spectrum for V2572 Sgr, RW UMi, CQ Vel and XX Tau for which radial velocities were measured from the H$\alpha$ emission line. The gray line marks the central lambda of H$\alpha$. The $\lambda$6678~{\AA} He{\sc i} emission line is also present in all spectra.}
\label{fig:spectra}
\end{figure*}
\noindent
%The individual light curves for the novae here are described. 

\begin{figure}
  \resizebox{0.5\textwidth}{!}{\includegraphics[trim={10mm 6.5mm 0mm 0mm}, clip]{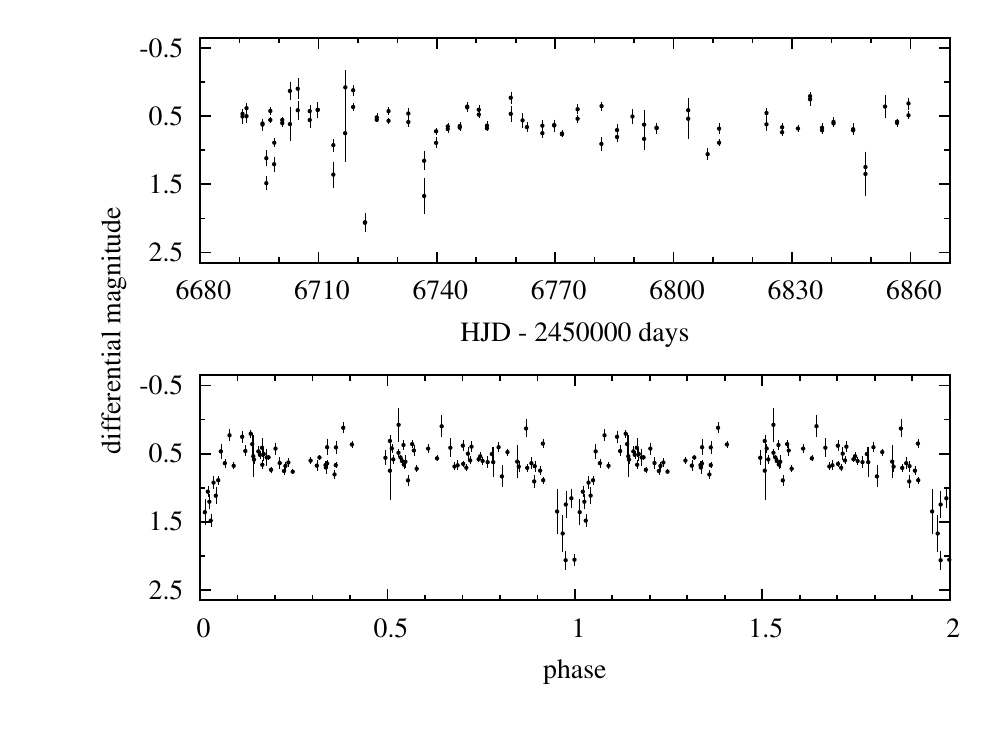}}
\caption{Top: The CTIO light curve of X Cir. Bottom: Phase light curve of this data according to ephemeris (\ref{eq:xcir}) described in section \ref{subs:xcir}.}
\label{fig:xcir-ctio}
\end{figure}

\begin{figure}
  \resizebox{0.5\textwidth}{!}{\includegraphics[trim={10mm 00mm 0mm 0mm}, clip]{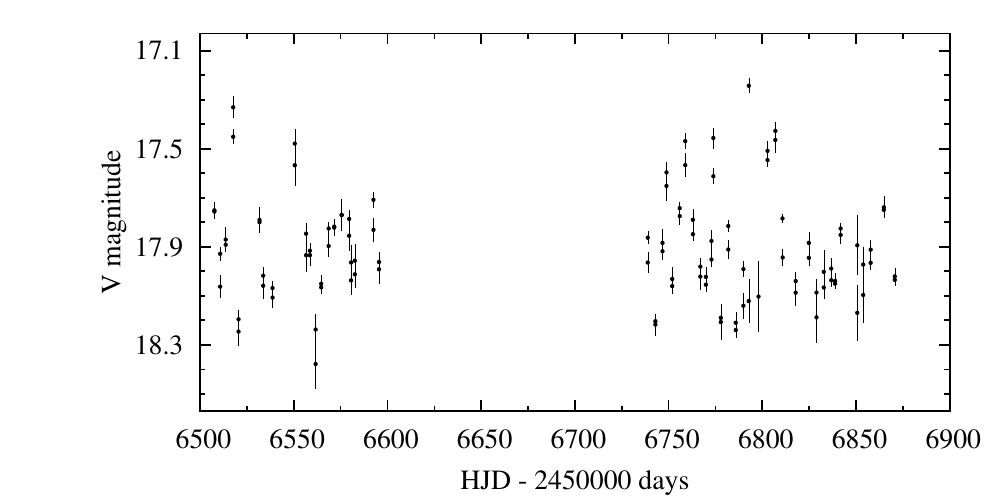}}
\caption{The CTIO light curves of V2572 Sgr described in section \ref{subs:v2572sgr}.}
\label{fig:v2572sgr-ctio}
\end{figure}

\begin{figure}
\centerline{\includegraphics[width=.5\textwidth]{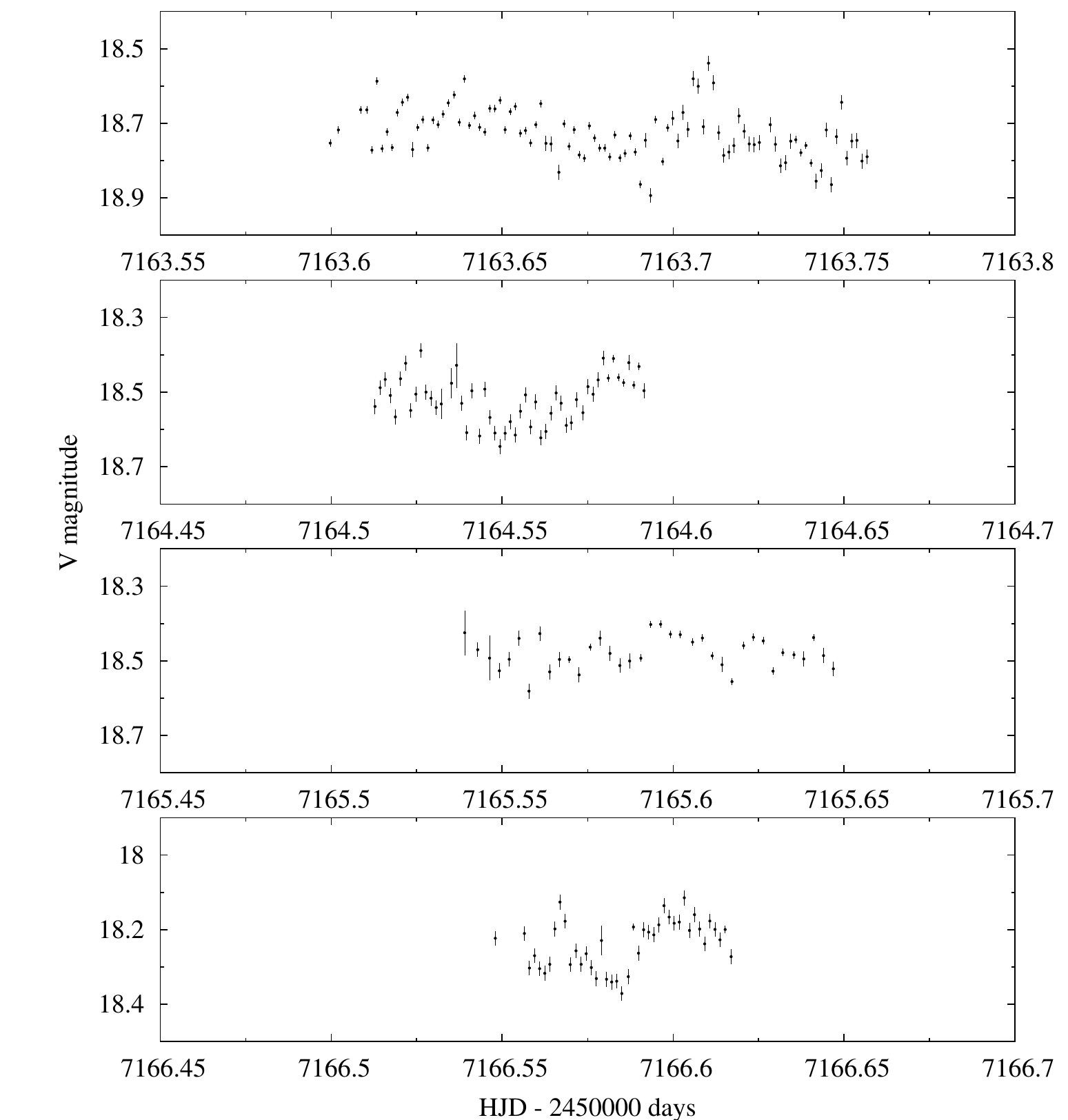}}
\caption{{\it V}-band light curves of IL Nor taken in 2015 at du Pont telescope. }
\label{fig:ilnor-lc}
\end{figure}

\begin{figure}
\centerline{\includegraphics[width=.5\textwidth]{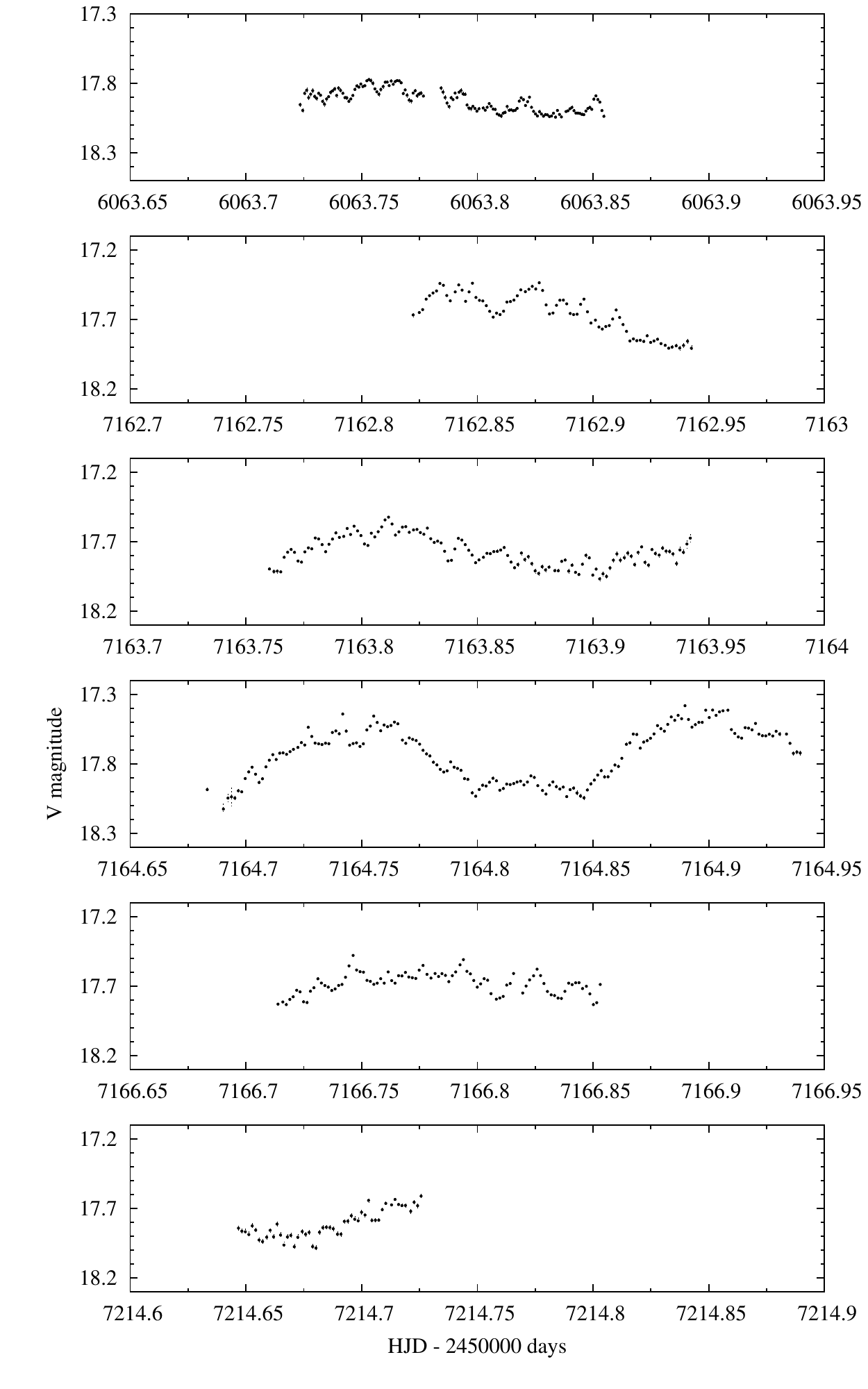}}
\caption{{\it V}-band Light curves of V2572 Sgr. The first one was observed with EFOSC2/NTT in May 2012 and the other ones with du Pont telescope in May-July 2015.}
\label{fig:v2572Sgr-lc}
\end{figure}

\begin{figure}
\centerline{\includegraphics[width=.5\textwidth]{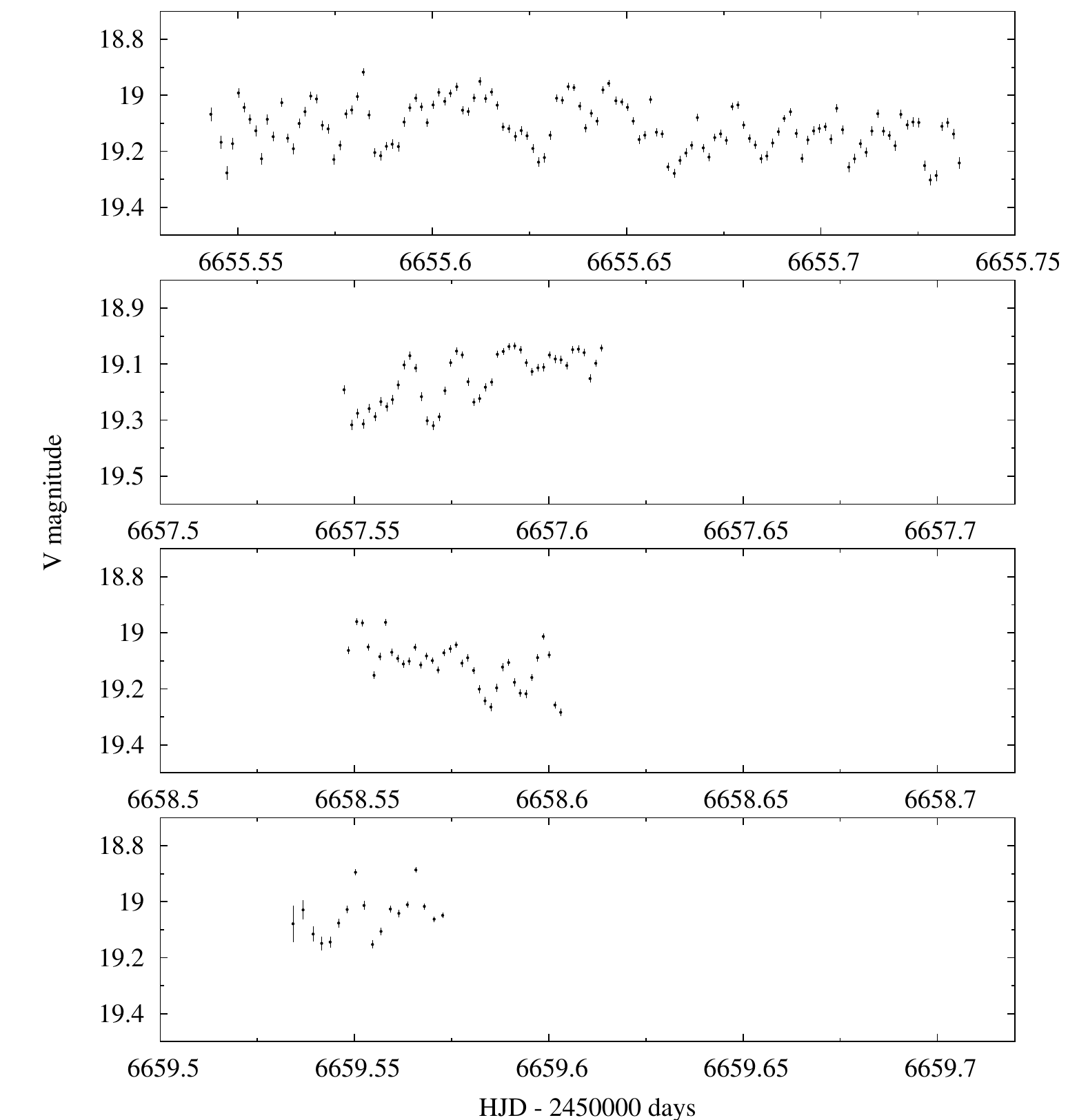}}
\caption{Light curves of XX Tau taken at du Pont telescope.}
\label{fig:xxtau-lc}
\end{figure}

\begin{figure}
\centerline{\includegraphics[width=.5\textwidth]{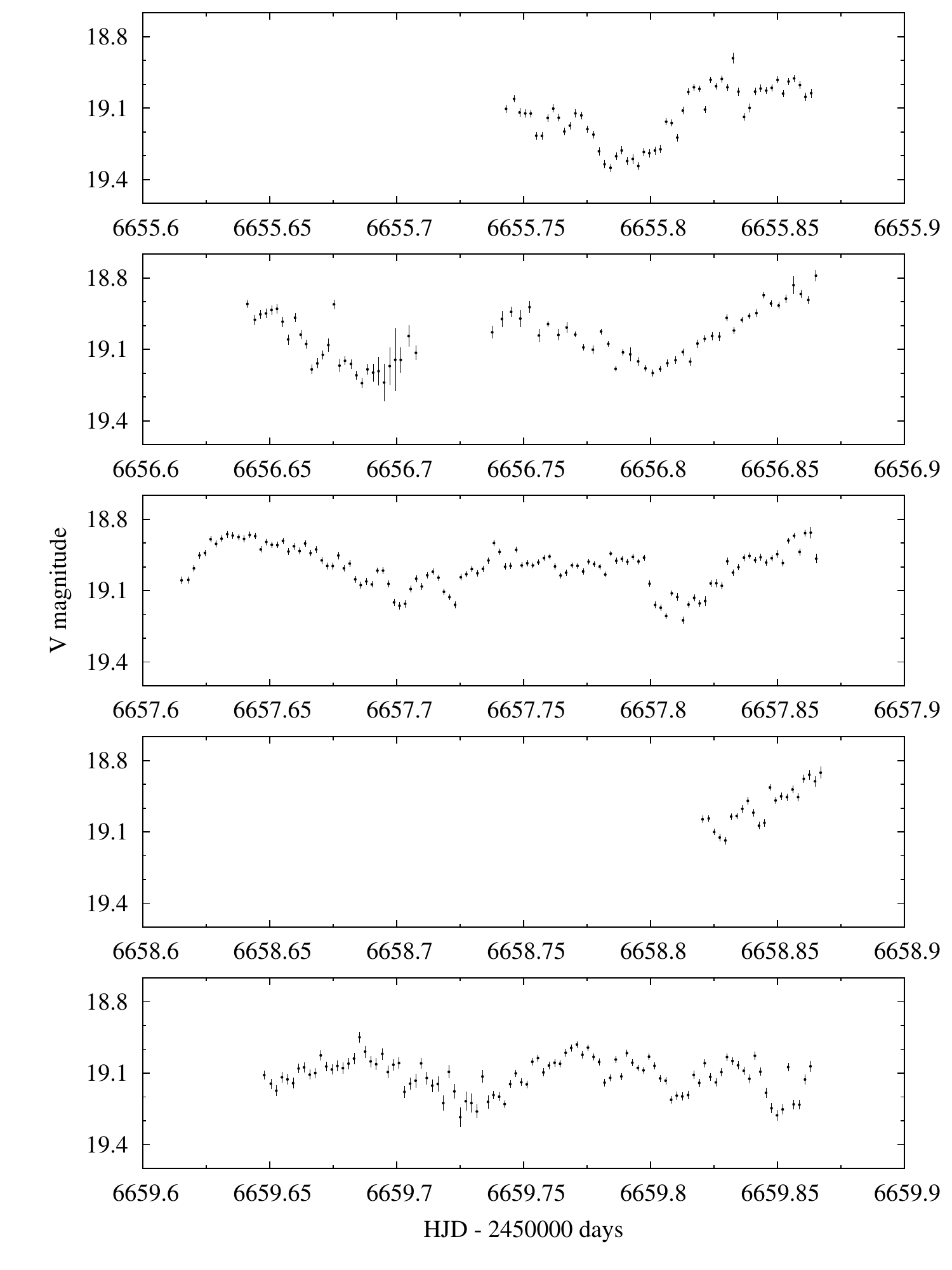}}
\caption{{\it V}-band Light curves of CQ Vel taken at du Pont telescope.}
\label{fig:cqvel-lc}
\end{figure}

%%%%%%%%%%%%%%%%%%%%%%%%%%%%%%%%%%%%%%%%
\begin{table}
\centering

%\resizebox{0.5\textwidth}{!}{\begin{minipage}{\textwidth}
\begin{minipage}{0.9\columnwidth}
 \caption{Eclipse epochs of the four eclipsing novae from the literature and from CTIO
observation data. The O-C values refer to the ephemerides given in Table \ref{tab:improvePorbs}.}  
%  \scalebox{0.9} {
\begin{tabular}{@{}lrrrl}
\hline
 \multicolumn{1}{l}{Object} &
 \multicolumn{1}{c}{E} &
   \multicolumn{1}{c}{$\mathrm{HJD}$} &
  \multicolumn{1}{c}{$\mathrm{O - C}$} &
   \multicolumn{1}{c}{Ref. } \\
   \multicolumn{1}{c}{ } &
  \multicolumn{1}{c}{ } &
 \multicolumn{1}{c}{$-2\,400\,000$ d} &
  \multicolumn{1}{c}{d}&
   \multicolumn{1}{c}{ } \\

\hline

WY Sge  & -14178  & 44881.639   &  0.0018  &  (1) \\
        & -14171  & 44882.711   & -0.0017  &  (1) \\
        & -14003  & 44908.524   &  0.0007  &  (1) \\
        & -12510  & 45137.8998  &  0.0001  &  (1) \\
        & -12498  & 45139.7434  &  0.0001  &  (1) \\
        &  11564  & 48836.4976  & -0.0001  &  (2) \\ 
        &  11571  & 48837.5726  & -0.0006  &  (2) \\
        &  14151  & 49233.9498  & -0.0005  &  (2) \\
        &  14157  & 49234.8722  &  0.0001  &  (2) \\
        &  14170  & 49236.869   & -0.0004  &  (2) \\
        &  14176  & 49237.791   & -0.0002  &  (2) \\
        &  14177  & 49237.945   &  0.0002  &  (2) \\
        &  63241  & 56775.8667  & -0.0035  &  (3) \\
        &  63306  & 56785.8580  &  0.0015  &  (3) \\
        &  63618  & 56833.7893  & -0.0012  &  (3) \\
        &  63637  & 56836.7094  & -0.0001  &  (3) \\
        &  63696  & 56845.7720  & -0.0020  &  (3) \\
        &  63884  & 56874.6585  &  0.0013  &  (3) \\
        &  63942  & 56883.5697  &  0.0017  &  (3) \\
        &  65552  & 57130.9224  &  0.0028  &  (3) \\

\hline

V728 Sco &  -14    &  56013.8704  &   0.0002   & (4)    \\
         &   -7    &  56014.8379  &  -0.0007   & (4)    \\
         &    0    &  56015.8073  &   0.0004   & (4)    \\
         &   29    &  56019.8182  &  -0.0005   & (4)    \\
         &  346    &  56063.6729  &   0.0008   & (4)    \\
         &  353    &  56064.6404  &  -0.0001   & (4)    \\
         &  354    &  56064.7750  &  -0.0010   & (4)    \\
         & 4988    &  56705.8401  &   0.0003   & (3)    \\
         & 5082    &  56718.8412  &  -0.0025   & (3)    \\
         & 5147    &  56727.8347  &  -0.0010   & (3)    \\
         & 5169    &  56730.8801  &   0.0010   & (3)    \\
         & 5234    &  56739.8729  &   0.0017   & (3)    \\
         & 5558    &  56784.6921  &  -0.0007   & (3)    \\
         & 6244    &  56879.5948  &   0.0017   & (3)    \\
         & 6446    &  56907.5382  &   0.0007   & (3)    \\
         & 6800    &  56956.5117  &   0.0023   & (3)    \\
         & 6865    &  56965.5034  &   0.0020   & (3)    \\
         & 7569    &  57062.8920  &   0.0002   & (3)    \\
         & 8154    &  57143.8197  &  -0.0001   & (3)    \\
         & 8709    &  57220.5947  &  -0.0031   & (3)    \\
         & 8804    &  57233.7386  &  -0.0014   & (3)    \\

\hline

OY Ara  &  -42829 &  49862.822     &  0.0001 &  (5) \\
        &       0 &  56516.5710    & -0.0012 &  (3) \\
        &     283 &  56560.5463    & -0.0009 &  (3) \\
        &    1160 &  56696.8211    & -0.0022 &  (3) \\
        &    1289 &  56716.8686    &  0.0002 &  (3) \\
        &    1366 &  56728.8375    &  0.0041 &  (3) \\

\hline

V849 Oph &     0  &   48799.7412   &  0.0000 &  (6)   \\
         &     1  &   48799.9149   &  0.0009 &  (6)   \\
         &   186  &   48831.8736   & -0.0003 &  (6)   \\
         &   191  &   48832.7384   &  0.0008 &  (6)   \\
         & 29884  &   53962.3846   & -0.0003 &  (7)   \\
         & 29890  &   53963.4197   & -0.0017 &  (7)   \\
         & 29895  &   53964.2840   & -0.0012 &  (7)   \\
         & 44687  &   56519.6881   & -0.0056 &  (3)   \\
         & 46487  &   56830.6621   &  0.0074 &  (3)   \\

\hline
\end{tabular}
 Ref:~~(1) \citet{shara84wysge}, (2) \citet{somers96wysge}, (3) this work, (4) \citet{tappertIII}, (5) \citet{zhao97oyar}, (6) \citet{shafter93DOaqlV849oph}, (7)  \citet{zengin2010v849oph} \\

%}
\label{tab:Ectio}
\end{minipage}
%\vspace{32.97583pt}
\end{table}

\begin{table}
\centering
 \caption{Spectroscopic measurements. The name of the nova, the radial velocities parameters defined in Eq. (\ref{eq:fit-rv}), the equivalent width for $H\alpha$ emission line and the radial velocity measures from that line together with its HJD are given.}

\begin{tabular}{@{}lr|lr}
\hline
   \multicolumn{1}{c}{$\mathrm{HJD}$} &
  \multicolumn{1}{c|}{$ \mathrm{v_r}$}  &
   \multicolumn{1}{c}{$\mathrm{HJD}$} &
    \multicolumn{1}{c}{$ \mathrm{v_r}$}  \\
  \multicolumn{1}{c}{$-2\,450\,000$ d} &
  \multicolumn{1}{c|}{(km/s)} &
  \multicolumn{1}{c}{$-2\,450\,000$ d} &
  \multicolumn{1}{c}{(km/s)} \\
  \hline   
   \multicolumn{2}{c|}{\textbf{V2572 Sgr}} & \multicolumn{2}{c}{\textbf{RW UMi}} \\

         5242.9278  &  68.52     &    7193.5948 & -158.45   \\ 
         5243.6482  &  60.60     &    7193.6020 & -166.56   \\ 
         5243.7272  &  10.86     &    7193.6092 & -138.88   \\ 
         5243.7956  & 116.31     &    7193.6164 & -147.71   \\ 
         5244.5174  &  26.86     &    7193.6237 & -143.23   \\ 
         5244.6005  &  71.65     &    7193.6309 & -137.75   \\ 
         5244.6470  &  34.06     &    7193.6381 & -140.61   \\ 
         5244.7140  &  56.87     &    7193.6453 & -146.05   \\ 
         5244.7475  &  73.38     &    7193.6525 & -173.64   \\ 
         5244.7789  &  23.29     &    7193.6597 & -162.63   \\ 
         5244.8102  &  42.36     &    7195.6015 & -142.06   \\ 
         5244.9086  &  36.79     &    7195.6087 & -149.77   \\ 
         5246.9286  &  -3.30     &    7195.6159 & -152.83   \\ 
 \multicolumn{2}{c|}{\textbf{XX Tau}}     &    7195.6231 & -144.41   \\ 
         8482.6026  & -31.96     &    7195.6303 & -148.72   \\ 
         8482.6100  & -16.86     &    7195.6376 & -129.93   \\ 
         8482.6174  &  30.55     &    7195.6448 & -135.47   \\ 
         8482.6247  & 222.31     &    7195.6520 & -144.77   \\ 
         8482.7278  &  16.99     &    7195.6592 & -114.02   \\ 
         8482.7352  &  45.82     &    7195.6664 & -134.63   \\ 
         8482.7425  & -49.28     &    7195.6736 & -161.44   \\ 
         8482.7499  & -41.51     &    7195.6809 & -168.36   \\ 
         8483.5690  &-140.80     &    7195.6881 & -157.43   \\ 
         8483.5764  &-175.68     &    7195.6953 & -136.91   \\ 
         8483.5837  &-239.57     &    7195.7025 & -152.90   \\ 
         8483.5911  &-110.32     &    7195.7097 & -135.46   \\ 
         8483.5992  & 109.02     &    7196.4224 & -125.13   \\ 
         8484.5368  &-133.65     &    7196.4296 & -135.27   \\ 
         8484.5441  & -15.55     &    7196.4368 & -138.73   \\ 
         8484.5515  &  43.39     &    7196.4440 & -147.87   \\ 
         8484.5589  &  40.82     &    7196.4512 & -161.69   \\ 
         8484.5667  &  33.30     &    \multicolumn{2}{c|}{\textbf{CQ Vel}}   \\ 
         8484.5741  &  67.47     &    6012.5655 &  90.20    \\ 
         8484.5815  & 125.17     &    6012.5958 & 142.19    \\ 
         8484.5888  & 133.73     &    6013.4878 &-143.14    \\ 
         8484.5967  & 162.77     &    6013.4991 &  74.73    \\ 
         8484.6040  & 206.05     &    6013.5101 &  42.50    \\ 
         8484.6114  & 205.86     &    6013.5212 & 112.31    \\ 
         8484.6188  &  84.78     &    6013.5356 & 179.21    \\ 
         8484.6265  &  47.56     &    6013.5466 & 100.48    \\ 
         8484.6339  &  -1.74     &    6013.5574 &  74.32    \\ 
         8484.6412  & -79.13     &    6013.5888 &  46.92    \\ 
         8484.6486  &-112.06     &    6013.5998 &  16.15    \\ 
         8484.6562  &-148.03     &    6014.4854 & 170.30    \\ 
         8484.6636  &-289.09     &    6014.4966 & -32.18    \\ 
         8484.6710  &-267.78     &    6014.5079 &  10.55    \\ 
         8484.6784  &-205.82     &    6014.5212 & 206.04    \\ 
         8524.5969  & -95.74     &    6014.5323 & 171.34    \\ 
         8524.6042  &-187.88     &    6014.5777 & 259.17    \\ 
         8524.6116  &-229.88     &    6014.6266 & -20.27    \\ 
         8524.6190  & -19.75     &    6015.4856 & 181.25    \\ 
         8526.5939  &  75.46     &    &\\
         8526.6013  &  86.15     &    &\\
         8526.6086  &  -4.80     &    &\\
         8526.6160  & -62.70     &    &\\ 
\hline

\end{tabular}
  %}

%\end{minipage}
\label{tab:rv}
\end{table}

%\begin{center}
\begin{table*}
\begin{minipage}{2.0\columnwidth}
\caption{ The orbital periods of old novae considered to create the distribution shown in Fig. \ref{fig:histograma}. The name, $P_\mathrm{{orb}}$, the method used to derived it and the references are presented (OM: photometric orbital modulation, RV: Radial velocity, E: eclipse and SH: superhump). Those novae with daily alias ambiguities are categorized as ``provisional'' and they are marked with $\ast$. The choice for the orbital period value presented here is discussed in section \ref{sec:summary}.  }
\label{tab:Porb}

 \begin{tabular}{lrrl|lrrl}
 \hline
%  &&&&&&&\\
  \multicolumn{1}{c}{Name} &
  \multicolumn{1}{c}{$P_\mathrm{{orb}}$(d)} &
  \multicolumn{1}{c}{method}  &
 \multicolumn{1}{l|}{reference} &
 \multicolumn{1}{|c}{Name} &
  \multicolumn{1}{c}{$P_\mathrm{{orb}}$(d)} &
  \multicolumn{1}{c}{method}  &
 \multicolumn{1}{l}{reference} \\
%  &&&&&&&\\
  
\hline

RW UMi      & 0.05912      & OM-RV    & (1), (2), This paper  & V2467 Cyg   & 0.1596       & OM       & (34)             \\        
GQ Mus      & 0.059365     & OM-RV    & (3), (4)              & DO Aql      & 0.167762     & E        & (35)             \\ 
CP Pup      & 0.061264     & OM-RV-SH & (5), (6)              & V849 Oph    & 0.17275611   & E        & (35), This paper \\
IL Nor$\ast$       & 0.06709      & OM       & This paper     & V697 Sco    & 0.187        & OM       & (36)             \\
V458 Vul    & 0.068126     & RV       & (7)                   & V825 Sco    & 0.19165877   & E        & (16)             \\
V1974 Cyg   & 0.08126      & OM-SH    & (8)                   & DQ Her      & 0.193621     & E-RV     & (37)             \\
RS Car$\ast$      & 0.082429     & OM-SH?   & (9), This paper & CT Ser      & 0.195        & RV       & (38)             \\ 
DD Cir      & 0.09746      & E        & (10)                  & AT Cnc      & 0.201634     & RV-OM    & (39), (40), (41) \\
V Per       & 0.107123     & E-RV     & (11)                  & T Aur       & 0.204378     & E        & (42)             \\
V597 Pup    & 0.11119      & E        & (12)                  & V446 Her    & 0.207        & RV       & (29)             \\
QU Vul      & 0.111765     & E        & (13)                  & V4745 Sgr   & 0.20782      & OM       & (43)             \\
CQ Vel$\ast$       & 0.11272      & OM-RV    & This paper     & HZ Pup      & 0.212        & RV       & (44)             \\
V2214 Oph   & 0.117515     & OM       & (14)                  & AP Cru      & 0.213        & OM       & (9)              \\
V630 Sgr    & 0.11793      & E-SH     & (15), (16)            & AR Cir      & 0.214        & OM-RV    & (22)             \\
V351 Pup    & 0.1182       & OM       & (15)                  & HR Del      & 0.214165     & RV       & (45)             \\
V5116 Sgr   & 0.1238       & E        & (16)                  & V5588 Sgr   & 0.214321     & OM       & (16)             \\
V4633 Sgr   & 0.1255667    & OM-SH    & (16), (17)            & NR TrA      & 0.219        & E-RV     & (46)             \\
V363 Sgr    & 0.126066     & OM       & This paper            & CN Vel      & 0.2202       & RV       & (22)             \\
DN Gem      & 0.127844     & RV       & (18), (19)            & V365 Car    & 0.223692     & OM-RV    & (22), This paper \\
V339 Del    & 0.1314       & OM       & (20)                  & V1039 Cen   & 0.247        & OM       & (47)             \\
V4742 Sgr   & 0.1336159    & E        & (16)                  & V1425 Aql   & 0.2558       & OM       & (48)             \\
V1494 Aql   & 0.134614     & E        & (21)                  & HS Pup      & 0.2671       & RV       & (22)             \\
V5585 Sgr   & 0.137526     & E        & (16)                  & V2615 Oph   & 0.272339     & OM       & (16)             \\
V603 Aql    & 0.138201     & OM-RV-SH & (18)                  & V4743 Sgr   & 0.2799       & OM       & (49)             \\
V728 Sco    & 0.13833866   & E-RV     & (22), This paper      & V972 Oph    & 0.281        & RV       & (22)             \\
V1668 Cyg   & 0.1384       & E        & (23)                  & BY Cir      & 0.2816       & E        & (10)             \\
XX Tau$\ast$& 0.13588       & RV       & This paper            & V2540 Oph   & 0.284781     & OM       & (50)             \\
DY Pup      & 0.13952      & E        & This paper            & V1059 Sgr   & 0.2861       & RV       & (51)             \\
V1500 Cyg   & 0.139613     & OM       & (24)                  & Z Cam       & 0.289841     & RV       & (52), (53)       \\
RR Cha      & 0.1401       & E-SH     & (9)                   & V959 Mon    & 0.29585      & OM       & (54)             \\
V909 Sgr    & 0.14286      & OM-RV    & (22)                  & V838 Her    & 0.297635     & E        & (55)             \\
RR Pic      & 0.145025959  & OM-SH    & (25), (26)            & V2275 Cyg   & 0.3145       & OM       & (56)             \\
CP Lac      & 0.145143     & RV       & (18)                  & BT Mon      & 0.333814     & E-RV     & (57)             \\
V500 Aql    & 0.1452       & OM-RV    & (27), (11)            & V2677 Oph   & 0.3443       & OM       & (16)             \\
V2468 Cyg   & 0.14525      & OM       & (28)                  & QZ Aur      & 0.357496     & E-RV     & (58), (59)       \\
V533 Her    & 0.147        & RV       & (29)                  & Q Cyg       & 0.42036      & RV       & (16)             \\
V2574 Oph   & 0.1477       & OM-SH    & (30)                  & J17014 4306 & 0.5340257    & E        & (60)             \\
V5113 Sgr   & 0.150015     & OM       & (16)                  & V841 Oph    & 0.601304     & RV       & (18)             \\
V4579 Sgr   & 0.15356146   & E        & (16)                  & V368 Aql    & 0.690509     & E        & (61)             \\
V992 Sco    & 0.15358      & OM       & (10)                  & V723 Cas    & 0.693277     & OM       & (62)             \\
V373 Sct    & 0.1536       & RV       & (22)                  & CP Cru      & 0.944        & E        & (10)             \\
WY Sge      & 0.153634547  & E        & (31), This paper      & V2674 Oph   & 1.30207      & E        & (16)             \\
X Cir       & 0.15445953   & E        & This paper            & X Ser       & 1.48         & RV       & (29)             \\
OY Ara      & 0.155390     & E-RV     & (32), This paper      & V5589 Sgr   & 1.5923       & E        & (16)             \\
V1493 Aql   & 0.156        & OM       & (33)                  & HV Cet      & 1.772        & OM       & (63)             \\ 
V2572 Sgr$\ast$    & 0.156215     & OM-RV    & This paper     & GK Per      & 1.996803     & RV       & (64)             \\

\hline

\end{tabular}

References:\\
(1) \citet{retter2001_rwumi}, (2) \citet{bianchini2003-RWUmi}, (3) \citet{diaz94GQmus}, (4) \citet{narloch2014GQmus}, (5) \citet{masonCPpup_phot}, (6) \citet{bianchiniCPpup_RV}, (7) \citet{rodriguez-gilV458vul}, (8) \citet{olechV1974cyg}, (9) \citet{woudt2002}, (10) \citet{woudt2003}, (11) \citet{haefnerVperV500Aql}, (12) \citet{warnerV597pup}, (13) \citet{shafter95QUvul}, (14)  \citet{baptista93V2214oph}, (15) \citet{woudt2001cqvel}, (16) \citet{mroz2015}, (17) \citet{Lipkin2008v4633sgr}, (18) \citet{peters2006}, (19) \citet{retter99DNgem}, (20) \citet{chocholv339del}, (21) \citet{katoV1494Aql}, (22) \citet{tappertIII}, (23)   \citet{kaluznyV1668cyg}, (24) \citet{pavlenkoV1500cyg}, (25) \citet{vogt2016}, (26) \citet{yo2018}, (27) \citet{haefner99V500aql}, (28) \citet{chocholV2468cyg}, (29) \citet{thorstensen2000}, (30) \citet{kang2006}, (31) \citet{somers96wysge}, (32) \citet{zhao97oyar}, (33) \citet{dobrotkaV1493aql}, (34) \citet{SwierczynskiV2467cyg}, (35) \citet{shafter93DOaqlV849oph}, (36) \citet{warnerV697sco}, (37) \citet{daiDQher2009}, (38) \citet{ringwaldCTser_v825Her}, (39) \citet{nogamiATcnc}, (40) \citet{shara2012atnc}, (41) \citet{ATcncperiod2019}, (42) \citet{daiTaur2010}, (43) \citet{dobrotkav4745sgr}, (44) \citet{hzpup2017thorstensen}, (45) \citet{kuersterHRdel}, (46) \citet{walterNRtra}, (47) \citet{woudt2005}, (48) \citet{retter98timescale}, (49) \citet{kang2006_V4743Sgr}, (50) \citet{akV2540oph}, (51) \citet{thorstensen2010}, (52) \citet{thorstensen1995}, (53) \citet{shara2007zcam}, (54) \citet{munariV959mon}, (55) \citet{ingramV838her}, (56) \citet{balmanv2275cyg}, (57) \citet{smithBTmon}, (58) \citet{szkody94QZaurRV}, (59) \citet{campbell95QZaurEclip}, (60) \citet{shara2017nature}, (61) \citet{marinV368aql}, (62) \citet{ochnerV723cas}, (63) \citet{beardmoreHVcet}, (64) \citet{morales2002gkper}

\end{minipage}
\end{table*}
%\end{center}

% Don't change these lines
\bsp	% typesetting comment
\label{lastpage}
\end{document}